\documentclass[aps,prl,twocolumn,groupedaddress,superscriptaddress,amsfonts,amssymb,amsmath,citeautoscript,a4paper]{revtex4-1}

\usepackage[utf8]{inputenc}
\usepackage[english]{babel}
\usepackage{microtype}
\usepackage{txfonts}
\usepackage{txfontsb}
\usepackage{bm}
\usepackage{xcolor}
\usepackage{graphicx}
\usepackage{float}
\usepackage{xspace}

\usepackage{enumerate}
\usepackage[inline]{enumitem}

\usepackage{siunitx}
\DeclareSIUnit[number-unit-product=]\percent{\char`\%} 

\usepackage[pdftex]{hyperref}
\hypersetup{colorlinks,
            linkcolor={blue!75!black!80!yellow},
            citecolor={blue!75!black!80!yellow},
            urlcolor={blue!75!black!80!yellow},
            pdfstartview=FitH}

\usepackage{etoolbox}
        
\newcommand{\ie}{i.e.\@\xspace}  
\newcommand{\cf}{cf.\@\xspace}
\newcommand{\eg}{e.g.\@\xspace}

\newcommand{\ep}{\epsilon}
\newcommand{\vep}{\varepsilon}
\newcommand{\pll}{\parallel}

\newcommand{\e}{\text{e}}
\newcommand{\iu}{\text{i}}
\let\Re\relax\DeclareMathOperator{\Re}{Re}
\let\Im\relax\DeclareMathOperator{\Im}{Im}

\newcommand{\tensorGrk}[1]{\overset{\text{$\leftrightarrow$}}{\bm{#1}}}

\begin{document}

\title{Harnessing Ultraconfined Graphene Plasmons to Probe the Electrodynamics of Superconductors}

\author{A.~T.~Costa}
\affiliation{\small International Iberian Nanotechnology Laboratory, 4715-330 Braga, Portugal}

\author{P.~A.~D.~Gon\c{c}alves}
\affiliation{\small Center for Nano Optics, University of Southern Denmark, DK-5230~Odense~M, Denmark}

\author{D.~N.~Basov}
\affiliation{\small Department of Physics, Columbia University, New York, NY~10027, USA}

\author{Frank~H.~L.~Koppens}
\affiliation{\small ICFO --- Institut de Ci\`{e}ncies Fot\`{o}niques, The Barcelona Institute of Science and Technology, 08860 Castelldefels (Barcelona), Spain}
\affiliation{\small ICREA --- Instituci{\'o} Catalana de Recera i Estudis Avan\c{c}ats, Barcelona, Spain}

\author{N.~Asger~Mortensen}
\affiliation{\small Center for Nano Optics, University of Southern Denmark, Campusvej 55, DK-5230~Odense~M, Denmark}
\affiliation{\small Danish Institute for Advanced Study, University of Southern Denmark, Campusvej 55, DK-5230~Odense~M, Denmark}
\affiliation{\small Center for Nanostructured Graphene, Technical University of Denmark, DK-2800 Kgs. Lyngby, Denmark}

\author{N.~M.~R.~Peres}
\affiliation{\small International Iberian Nanotechnology Laboratory, 
4715-330 Braga, Portugal}
\affiliation{\small Centro de F\'{i}sica das Universidades do Minho e do Porto, Universidade do Minho, Campus de Gualtar, 4710-057 Braga, Portugal}
\affiliation{\small Departamento de 
F\'{i}sica and QuantaLab, Universidade do Minho, Campus de Gualtar, 4710-057 Braga, Portugal}

\begin{abstract}
We show that the Higgs mode of a superconductor, which is usually challenging to observe by far-field optics, can be made clearly visible using near-field optics by harnessing ultraconfined graphene plasmons. As near-field sources we investigate two examples: graphene plasmons and quantum emitters. In both cases the coupling to the Higgs mode is clearly visible. In the case of the graphene plasmons, the coupling is signaled by a clear anticrossing stemming from the interaction of graphene plasmons with the Higgs mode of the superconductor. In the case of the quantum emitters, the Higgs mode is observable through the Purcell effect. When combining the superconductor, graphene, and the quantum emitters, a number of experimental knobs become available for unveiling and studying the electrodynamics of superconductors. 
\end{abstract}

\maketitle



The superconducting state is characterized by a spontaneously broken continuous symmetry~\cite{Anderson1963}. As a consequence of the Nambu--Goldstone theorem, superconductors are expected to display two kinds of elementary excitations: the so-called Nambu--Goldstone (NG) and Higgs modes~\cite{Shimano2020,Pekker2015,Anderson2015}. The NG mode is associated with fluctuations of the phase of the order parameter, whereas the Higgs mode is related to amplitude fluctuations of the same. In superconductors and electrically charged plasmas, the NG (phase) mode couples to the electromagnetic field and its spectrum effectively acquires a gap (mass) due to the long-range Coulomb interaction (Anderson--Higgs mechanism)~\cite{Shimano2020}; this gap corresponds to the system's plasma frequency~\cite{Anderson1963,Higgs1964a,Higgs1964b}. On the other hand, the Higgs (amplitude) mode is always gapped, and in superconductors its minimum energy is equal to twice the superconducting gap~\cite{Littlewood1982}. 
Curiously, one often encounters in the literature statements that the Higgs mode does not couple to electromagnetic fields in linear response, making it difficult to observe in optical experiments~\cite{Shimano2020,Yang2019}. Experimental detection has only been achieved through higher-order response~\cite{CommentHiggsDetection}, \eg, by pumping the superconductor with intense terahertz (THz) fields and measuring the resulting oscillations in the superfluid density~\cite{Matsunaga2012,Matsunaga2013,Matsunaga2014,Matsunaga2017,Katsumi2018}. 

Naturally, the light--Higgs coupling is subjected to conservation laws, whereby translational invariance manifests in the conservation of wave vectors. Since far-field photons carry little momentum, wave vector conservation cannot be satisfied and the coupling is suppressed. However, little attention has been given to the fact that, strictly speaking, the linear-response coupling of the electromagnetic field to the Higgs mode only effectively vanishes in the $q \to 0$ limit~\cite{Yang2019,sun2020}. As such, at finite wave vectors---\ie, in the nonlocal regime---, the linear optical conductivity of the superconductor yields a finite contribution associated with the coupling to the Higgs mode~\cite{rickayzen1965theory,Yang2019,sun2020}. Hence, electromagnetic near-fields provided by, for instance, plasmons, emitters, or small scatterers, can couple to such amplitude fluctuations and therefore constitute a feasible, promising avenue toward experimental observations of the Higgs mode in superconductors. 
In this context, ultraconfined graphene plasmons~\cite{GoncalvesPeresBook,Goncalves_SpringerTheses} constitute a new paradigm for probing quantum nonlocal phenomena in nearby metals~\cite{Lundeberg:2017,Iranzo2018,Dias2018,Goncalves:2020a,Goncalves_SpringerTheses,Goncalves:2020b}, while their potential as tools for studying the intriguing electrodynamics of strongly correlated matter~\cite{Basov:2005,Basov:2011,Bouscher2017} remains largely virgin territory.

\begin{figure}[hb]
    \centering
    \includegraphics[width=1.0\columnwidth]{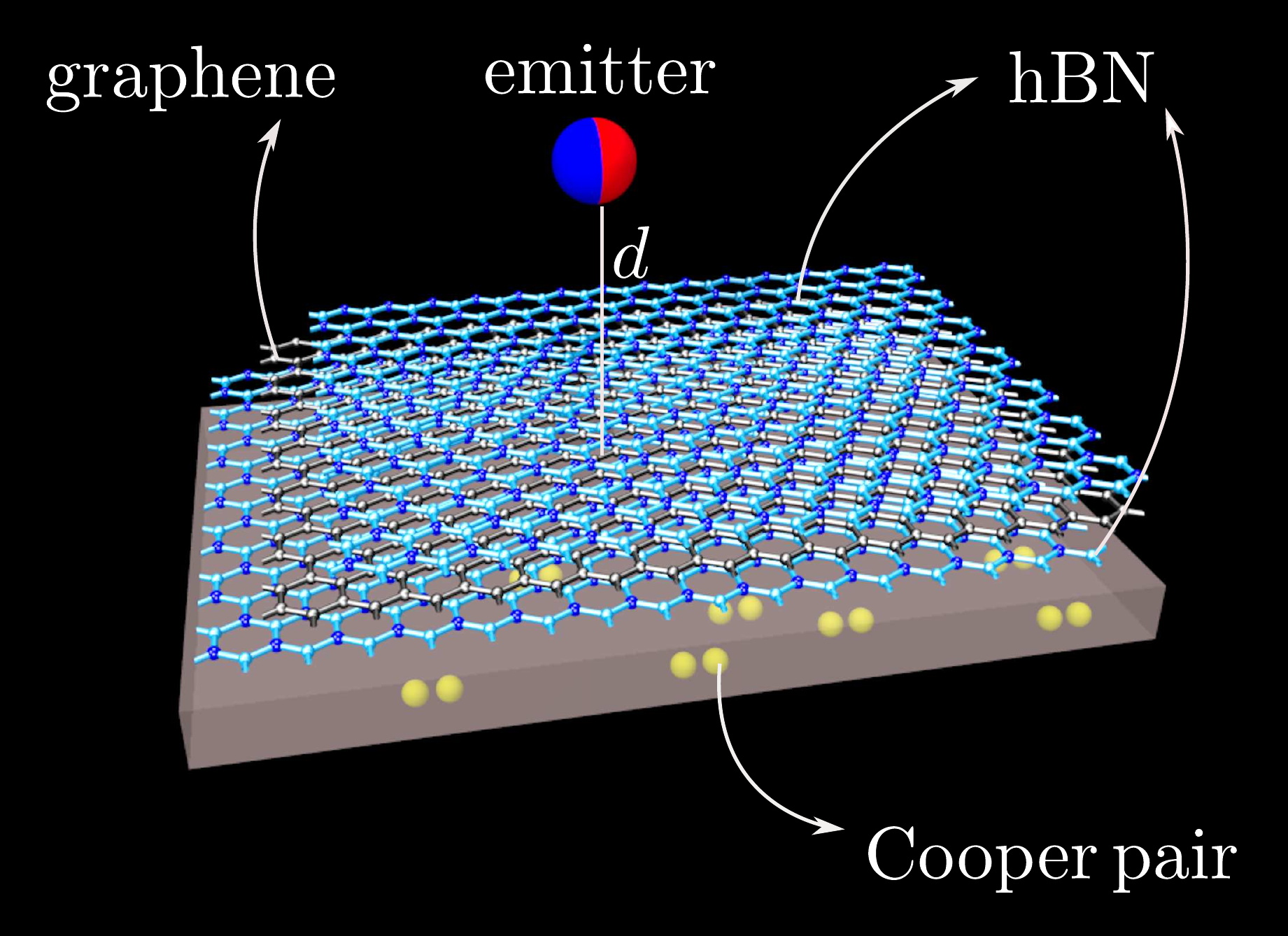}
    \caption{Schematic of the graphene--superconductor hybrid device considered here. Illustration of the heterostructure composed of a superconducting substrate, a few atomic layers of hexagonal boron nitride (hBN), a single sheet of graphene and a capping layer of hBN.  It should be noted that although here the hBN has been depicted in monolayer form, but our model can accommodate any number of hBN layers. The red-blue sphere represents an electric dipole placed above the heterostructure.}
    \label{fig:scheme}
\end{figure}

Here, we exploit the unprecedented field confinement yielded by graphene plasmons (GPs)~\cite{Iranzo2018,Dias2018,Goncalves:2020a,Goncalves_SpringerTheses,Goncalves:2020b,Epstein:2020} for investigating the near-field electromagnetic response of a heterostructure composed of a graphene sheet separated from a superconductor by a thin dielectric slab (see Fig.~\ref{fig:scheme}). Both the superconductor and the graphene sheet are characterized by their optical conductivity tensors~\cite{rickayzen1965theory,GoncalvesPeresBook}. The optical conductivity tensor of the superconductor is intrinsically nonlocal~\cite{rickayzen1965theory}, whereas for graphene it is possible to employ a local-response approximation at wave vectors much smaller than graphene's Fermi wave vector~\cite{GoncalvesPeresBook,Iranzo2018,Dias2018}. We show that the coupling between the Higgs mode in the superconductor and plasmons in the graphene manifest itself through the existence of an anticrossing-like feature in the near-field reflection coefficient. Furthermore, the energy and wave vector associated with this feature can be continuously tuned using multiple knobs, \eg, by changing 
(i) the temperature of the superconductor, 
(ii) the Fermi level of the graphene sheet, or
(iii) the graphene--superconductor separation. 
%
Finally, we suggest an alternative observation of the GPs-Higgs coupling through the measurement of the Purcell enhancement~\cite{Purcell:1946,Novotny_book,Goncalves_SpringerTheses} near the heterostructure. To that end, we calculate the electromagnetic local density of states (LDOS) above the graphene--dielectric--superconductor heterostructure; our results show that, in the absence of graphene, the coupling between the superconductor's surface polariton and its Higgs mode leads to an enhancement of the LDOS near the frequency of the latter. The presence of graphene changes qualitatively the behavior of the decay rate around the frequency of the Higgs mode, depending strongly on the emitter--graphene distance.


\section*{Theoretical background}

\subsection*{Electrodynamics of BCS-like superconductors}

The electrodynamics of superconductors and other strongly correlated matter constitutes a fertile research area~\cite{Basov:2005,Basov:2011}. In the following, we assume that the superconducting material is well-described by the Bardeen--Cooper--Schrieffer (BCS) theory of superconductivity~\cite{rickayzen1965theory,Bardeen:1957,Schrieffer_SCbook}. Chiefly, the microscopically derived linear optical conductivity tensor of a superconductor requires a nonlocal framework due to the finiteness of the Cooper-pair wave function. For homogeneous superconducting media, the longitudinal and transverse components of the nonlocal optical conductivity tensor---while treating nonlocality to leading-order~\cite{sigmaHomogeneous}---can be expressed as~\cite{Keller1990,Keller:1991,rickayzen1965theory}
\begin{subequations}
 \begin{align}
 \sigma_\text{L}(q,\omega) &= \sigma_{\text{D}}(\omega)   
 \frac{1}{ 1 - 3 \bar{\alpha}(\omega,T) \left( \frac{q c }{\omega} \right)^2 } 
 , \label{eq:sigmaL} \\[0.25em]
  \sigma_\text{T}( q,\omega) &= \sigma_{\text{D}}(\omega) 
  \left[ 1 + \bar{\alpha}(\omega,T) \left( \frac{q c }{\omega} \right)^2 \right] 
  , \label{eq:sigmaT} 
 \end{align}
respectively, where $\sigma_{\text{D}}(\omega) = \frac{\iu n e^2}{m(\omega + \iu \gamma)} $ is the Drude-like conductivity, and the dimensionless coefficient $\bar{\alpha}(\omega,T)$ amounts to 
\begin{align}
 \bar{\alpha}(\omega,T) = &\frac{\hbar^4}{30 \pi^2 n m^3 c^2} 
 \int_{0}^{\infty} \mathrm{d}k \, k^6 \nonumber\\
 &\times \left\{
 \frac{ 2 f(E_{\mathbf{k}}) [ 1 - f(E_{\mathbf{k}}) ]}{k_\text{B} T}
\left[1-\frac{\Delta_0^2(T)}{E_{\mathbf{k}}^2}\right] \right. \nonumber\\
 &+ \left.  \frac{(\hbar\omega)^2 \Delta_0^2(T)}{E_{\mathbf{k}}^3}
\frac{1-2f(E_{\mathbf{k}})}{(\hbar\omega)^2-(2E_{\mathbf{k}})^2}
 \right\}
. \label{eq:defalpha}
\end{align}%
\label{eqs:sigmaSC}%
\end{subequations}%
In the previous expression, ${E_{\mathbf{k}} = \sqrt{ ( \vep_{\mathbf{k}} - \mu)^2 + \Delta_{0}^2(T) } }$ is the quasiparticle excitation energy at temperature $T$, where ${\mu \simeq E_{\text{F}} = \frac{\hbar^2}{2m} (3 \pi^2 n)^{2/3}}$ is the superconductor's chemical potential, ${\vep_{\mathbf{k}} = \hbar^2k^2/2m}$ is the single-particle energy of an electron with wave vector $\mathbf{k}$, ${ \Delta_{0}(T) \equiv \Delta_{\mathbf{k} \to 0}(T)  = 1.76\times k_{\text{B}} T_c [1 - (T/T_c)^4]^{1/2} \Theta(T_c - T)}$ is the temperature-dependent gap parameter of the superconductor, and ${f(E_{\mathbf{k}}) = \left[ \exp(E_{\mathbf{k}}/k_{\text{B}} T) + 1 \right]^{-1}}$ is the Fermi--Dirac distribution. 

In possession of the response functions epitomized by Eqs.~(\ref{eqs:sigmaSC}), we employ the semiclassical infinite barrier (SCIB) formalism~\cite{FordWeber:1984,Goncalves_SpringerTheses} to describe electromagnetic phenomena at a planar dielectric--superconductor interface~\cite{Keller:1991,Keller1990,Keller1989}. Within this framework, the corresponding reflection coefficient for $p$-polarized waves is given by~(see SI Appendix)~\cite{FordWeber:1984,Goncalves_SpringerTheses}
\begin{subequations}
 \begin{equation}
  r_p^{\text{\textsc{sc}}} = \frac{ k_{z,\text{d}} - \ep_{\text{d}}\, \Xi }{k_{z,\text{d}} + \ep_{\text{d}}\, \Xi  } 
  , \label{eq:r_p_D-SC} 
 \end{equation}
with $k_{z,\text{d}} = \sqrt{\ep_{\text{d}}\frac{\omega^2}{c^2} - q_\pll^2}$, and $\Xi$ has the form
\begin{equation}
  \Xi = \frac{\iu}{\pi} \int_{-\infty}^{\infty} \frac{\mathrm{d} q_\perp}{q^2} 
  \left[\frac{q_\pll^2}{\ep_{\text{L}}(q,\omega) \vphantom{\left(\frac{q c}{\omega}\right)^2} } 
  + \frac{q_\perp^2}{ \ep_{\text{T}}(q,\omega) - \left(\frac{q c}{\omega}\right)^2 }\right] 
  , \label{eq:Xi_SC}
 \end{equation}%
\label{eqs:r_p_D-SC_interface}%
\end{subequations}%
where $q^2 = q_\pll^2 + q_\perp^2$, and $ \ep_{\text{L,T}} = \ep_\infty + \iu \sigma_{\text{L,T}}/(\omega \ep_0)$ are the components of the superconductor's nonlocal dielectric tensor (we take $\ep_\infty = 1$ hereafter). 

In what follows, we assume a typical high-$T_c$ superconductor, such as yttrium barium copper oxide (YBCO), with a normal state electron density of $n = \SI{6}{\per\cubic\nm}$ and a transition temperature of $T_c = \SI{93}{\kelvin}$ (yielding a superconducting gap of $\Delta_0 (0) \approx \SI{14.2}{\meV}$)~\cite{Keller:1991,Keller1990,PhysPropHighTcBook}.

\subsection*{Electrodynamics in graphene--dielectric--superconductor heterostructures}

With knowledge of the reflection coefficient for the dielectric--superconductor interface~(\ref{eqs:r_p_D-SC_interface}), the overall reflection coefficient, \ie, that associated with the dielectric--graphene--dielectric--superconductor heterostructure, follows from imposing Maxwell's boundary conditions~\cite{Jackson} at all the interfaces that make up the layered system. At the interface defined by the two-dimensional graphene sheet, the presence of graphene enters via a surface current with a corresponding surface conductivity~\cite{GoncalvesPeresBook}.

\begin{figure*}[t]
 \centering
  \includegraphics[width=1.0\textwidth]{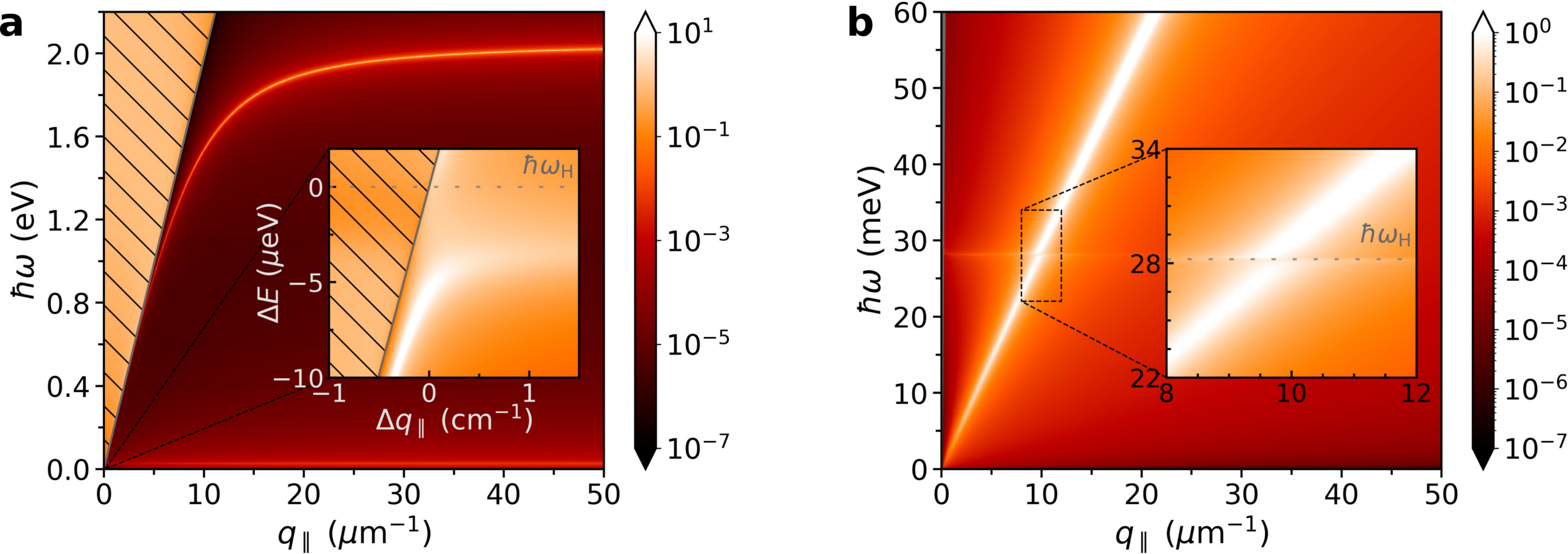}
  \caption{Spectra of surface electromagnetic waves in superconductors (\textbf{a}) and graphene--superconductor (\textbf{b}) structures, obtained from the calculation of the corresponding $\Im r_p$. 
  \textbf{a} Dispersion diagram of SPPs supported by a vacuum--superconductor interface (the hatched area indicates the light-cone in vacuum). The inset shows a close-up of an \emph{extremely} small region (notice the change of scale) where the SPP dispersion crosses the energy associated with the superconductor's Higgs mode; here, $\Delta E = E - \hbar\omega_{\text{H}}$ and $\Delta q_\pll = q_\pll  - \omega_{\text{H}}/c$.
  \textbf{b} Dispersion relation of GPs exhibiting an anticrossing feature that signals their interaction with the Higgs mode of the nearby superconductor; the graphene--superconductor separation is $t = \SI{5}{\nm}$. 
  Setup parameters: We take $T = \SI{1}{\kelvin}$; moreover, $n = \SI{6e21}{\per\cubic\centi\meter}$ (so that $E_{\text{F}} \approx \SI{1.20}{\eV} $ and $\hbar \omega_{\text{p}} \approx \SI{2.88}{\eV} $), $\hbar\gamma = \SI{1}{\micro\eV}$, and $T_c = \SI{93}{\kelvin}$ for the superconductor~\cite{Keller1989,Keller:1991,PhysPropHighTcBook}, and $E_{\text{F}}^{\text{gr}} = \SI{0.3}{\eV}$ and $\hbar\gamma^{\text{gr}} = \SI{1}{\meV}$, for graphene's Drude-like optical conductivity~\cite{Ni:2018}.
  }\label{fig:SC_vs_G-SC}
\end{figure*}

Signatures of system's collective excitations can then be found by analyzing the poles of corresponding reflection coefficient, which are identifiable as features in the imaginary part of the (overall) reflection coefficient, $\Im r_p$~(see  SI Appendix).

\section*{Coupling of the Higgs mode of a superconductor with graphene plasmons}

\subsection*{Signatures of the Higgs mode probed by graphene plasmons}
\label{ssec:G-SC}

Like ordinary conductors~\cite{Stockman:2018}, superconductors can also sustain surface plasmon polaritons (SPPs)~\cite{Stinson:2014,Basov:2016}. In turn, these collective excitations can couple to the superconductor's Higgs mode~\cite{Keller:1991,Keller1990}. Typically such interaction is extremely weak due to the large mismatch between superconductor's plasma frequency, $\omega_{\text{p}}$, and that of its Higgs mode, $\omega_{\text{H}} = 2\Delta_0/\hbar$; for instance, $\omega_{\text{H}}/\omega_{\text{p}} \sim 10^{-2}$, with $\omega_{\text{p}}$ and $\omega_{\text{H}}$ falling, respectively, in the visible and THz spectral ranges. As a result, at frequencies around $\omega_{\text{H}}$ the SPP resembles light in free-space and thus the SPP-Higgs coupling is essentially as weak as when using far-field optics (Fig.~\ref{fig:SC_vs_G-SC}a).

On the other hand, graphene plasmons not only span the THz regime but also attain sizable plasmon wave vectors at those frequencies~\cite{GoncalvesPeresBook,Goncalves_SpringerTheses}. Moreover, when the graphene sheet is near a metal---or a superconductor for that matter---graphene's plasmons become screened and acquire a nearly linear (acoustic) dispersion, pushing their spectrum further toward lower frequencies (\ie, a few THz) and larger wave vectors~\cite{Lundeberg:2017,Iranzo2018,Dias2018,Goncalves_SpringerTheses,Goncalves:2020a,Epstein:2020}. Therefore, these properties of acoustic-like GPs can be harnessed by placing a graphene monolayer near a superconducting surface, thereby allowing the interaction of graphene's plasmons with the Higgs mode of the underlying superconductor (Fig.~\ref{fig:SC_vs_G-SC}b). In this case the plasmon-Higgs interaction is substantially enhanced, a fact that is reflected in the observation of a clear anticrossing in the GP's dispersion near $\omega_{\text{H}}$, which, crucially, is orders of magnitude larger than that observed in the absence of graphene (\cf~Fig.~\ref{fig:SC_vs_G-SC}a--b).

\begin{figure*}[t]
 \centering
  \includegraphics[width=0.9\textwidth]{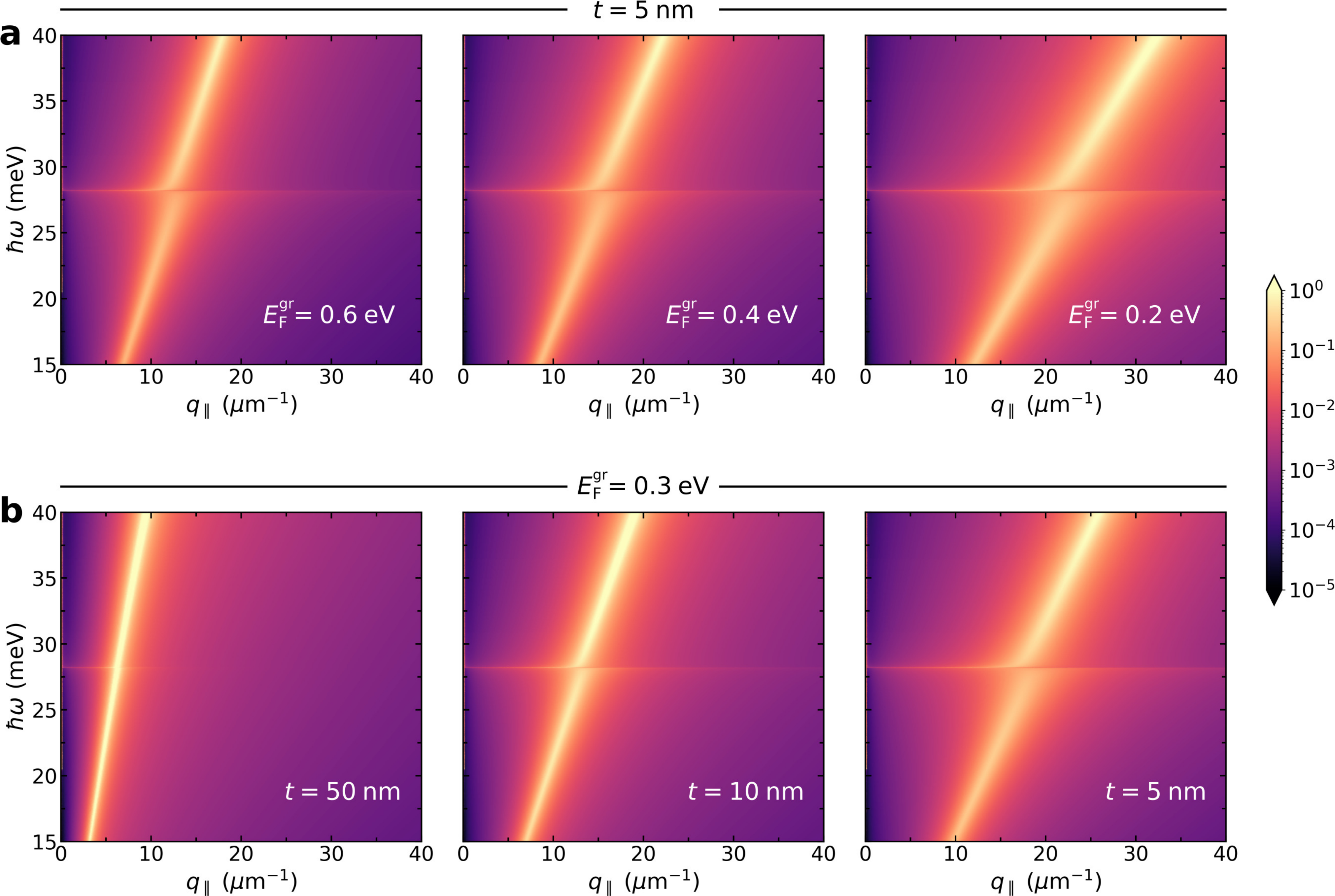}
  \caption{Tuning the hybridization of acoustic-like plasmons in graphene with the Higgs mode of a superconductor in air--hBN--graphene--hBN--superconductor heterostructures. The colormap indicates the loss function via $\Im r_p$.
  Spectral dependence upon varying the Fermi energy of graphene (\textbf{a}) and the graphene--superconductor distance (\textbf{b}).
  Setup parameters: the parameters of the superconductor are the same as in Fig.~\ref{fig:SC_vs_G-SC}, and the same goes for graphene's Drude damping. The thickness of the bottom hBN slab is given by $t$, whereas the thickness of the top hBN slab, $t'$, has been kept constant ($t'=\SI{10}{\nm}$). Here, we have modeled hBN's optical properties using a dielectric tensor of the form $\tensorGrk{\ep}_{\text{hBN}} = \text{diag}[\ep_{xx}, \ep_{yy}, \ep_{zz}]$ with $\ep_{xx} = \ep_{yy} = 6.7$ and $\ep_{zz} = 3.6$~\cite{Cai:2007,Woessner2015,Lundeberg:2017}.
  }\label{fig:G-SC_varying_EfGr-t}
\end{figure*}

Furthermore, the use of graphene plasmons for probing the superconductor's Higgs mode comes with the added benefit of control over the plasmon-Higgs coupling by tuning graphene's Fermi energy electrostatically~\cite{GoncalvesPeresBook,Goncalves_SpringerTheses,Chen:2012,Fei2012,Woessner2015}. This is explicitly shown in Fig.~\ref
{fig:G-SC_varying_EfGr-t}a, for a vacuum--hBN--graphene--hBN--superconductor heterostructure; as before, the coupling of GPs with the superconductor's Higgs mode manifests itself through the appearance of an avoided crossing in the vicinity of $\omega_{\text{H}}$, which occurs at successively larger wave vectors upon decreasing $E_{\text{F}}^{\text{gr}}$. 
Another source of tunability is the graphene--superconductor distance, $t$ (which, in the present configuration, corresponds to the thickness of the bottommost hBN slab). Strikingly, current experimental capabilities allow the latter to be controlled with atomic precision~\cite{Lundeberg:2017,Iranzo2018,Epstein:2020}. We exploit this fact in Fig.~\ref{fig:G-SC_varying_EfGr-t}b, where we have considered the same heterostructure, but now we have varied $t$ instead, while keeping $E_{\text{F}}^{\text{gr}}$ fixed. Naturally, the manifestation of the GP--Higgs mode interaction seems to be more pronounced for smaller $t$, reducing to faint feature at large $t$ (\cf the result for $t = \SI{50}{\nm}$). 
Lastly, it should be noted that the net effect of decreasing the graphene--superconductor separation $t$ is the outcome of two intertwined contributions: the graphene--superconductor interaction is evidently stronger when the materials lie close together, but equally important is the fact that the (group) velocity of plasmons in the graphene sheet gets continuously reduced as $t$ diminishes due to the screening exercised by the nearby superconductor (and, consequently, the GP's dispersion shifts toward higher wave vectors, eventually reaching the nonlocal regime~\cite{Goncalves_SpringerTheses,Lundeberg:2017,Goncalves:2020a}).


\subsection*{Higgs mode visibility through the Purcell effect}
\label{sec:Purcell}

One way to overcome the momentum mismatch and investigate the presence of electromagnetic surface modes is to place a quantum emitter~\cite{GoncalvesPeresBook,Koppens2011,Schadler2019,Kurman2018} (herein modeled as a point-like electric dipole) in the proximity of an interface and study its decay rate as a function of the emitter--surface distance. With the advent of atomically thin materials, and hBN in particular, all the relevant distances, \ie, emitter--superconductor, emitter--graphene, and graphene--superconductor, can be tailored with nanometric precision (\eg, by controlling the number of stacked hBN layers (each $\sim \SI{0.7}{\nm}$-thick)~\cite{Iranzo2018,Epstein:2020} or using atomic layer deposition~\cite{Scarafagio:2019,Cano:2020}). Although the availability of good emitters in the THz range are unarguably limited, semiconductor quantum dots with intersublevel transitions in this range and the relatively long relaxation times do exist~\cite{Zibik2009}. The modification of the spontaneous decay rate of an emitter is a repercussion of a change in the electromagnetic LDOS, $\rho(\mathbf{r})$, and it is known as the Purcell effect~\cite{Purcell:1946,Novotny_book,Goncalves_SpringerTheses}. Specifically, the Purcell factor---defined as the ratio $\frac{\rho(\mathbf{r})}{\rho_0(\mathbf{r})}$, where $\rho_0(\mathbf{r})$ is the LDOS experienced by an emitter in free-space---can be greatly enhanced by positioning the emitter near material interfaces supporting electromagnetic modes (which are responsible for augmenting the LDOS). In passing, we note that this LDOS enhancement does not strictly require an ``emitter'', since it can also be probed through the interaction of the sample with the illuminated tip of a near-field optical microscope (which may be modeled as an electric dipole in a first approximation)---in fact, most tip-enhanced spectroscopies rely on this principle~\cite{Schmid:2013,Deckert:2017,Chen:2019}.

Since in the near-field region the overall LDOS is dominated by contributions from $p$-polarization (and since plasmons possess $p$-polarization), in the following we neglect $s$-polarization contributions coming from the scattered fields. Then,
the orientation-averaged Purcell factor---or, equivalently, the LDOS enhancement---can be determined via~\cite{Novotny_book}
\begin{equation}
\frac{\rho(z)}{\rho_0} =  1 + \frac{1}{2} \int_0^\infty \mathrm{d}s \Re\left[\left(\frac{s^3}{s_z} -s s_z\right) r_p \, \e^{2 \iu \frac{\omega}{c} z s_z}\right] ,
\label{eq:rho}
\end{equation}
where $s_z=\sqrt{1-s^2}$, with $s = q_\parallel c/\omega$ denoting a dimensionless in-plane wave vector, $z = d - t'$ is the vertical coordinate relative to the surface of the topmost hBN layer, and where $d$ is the emitter--graphene distance.

\begin{figure}[t]
    \centering
    \includegraphics[width=\columnwidth]{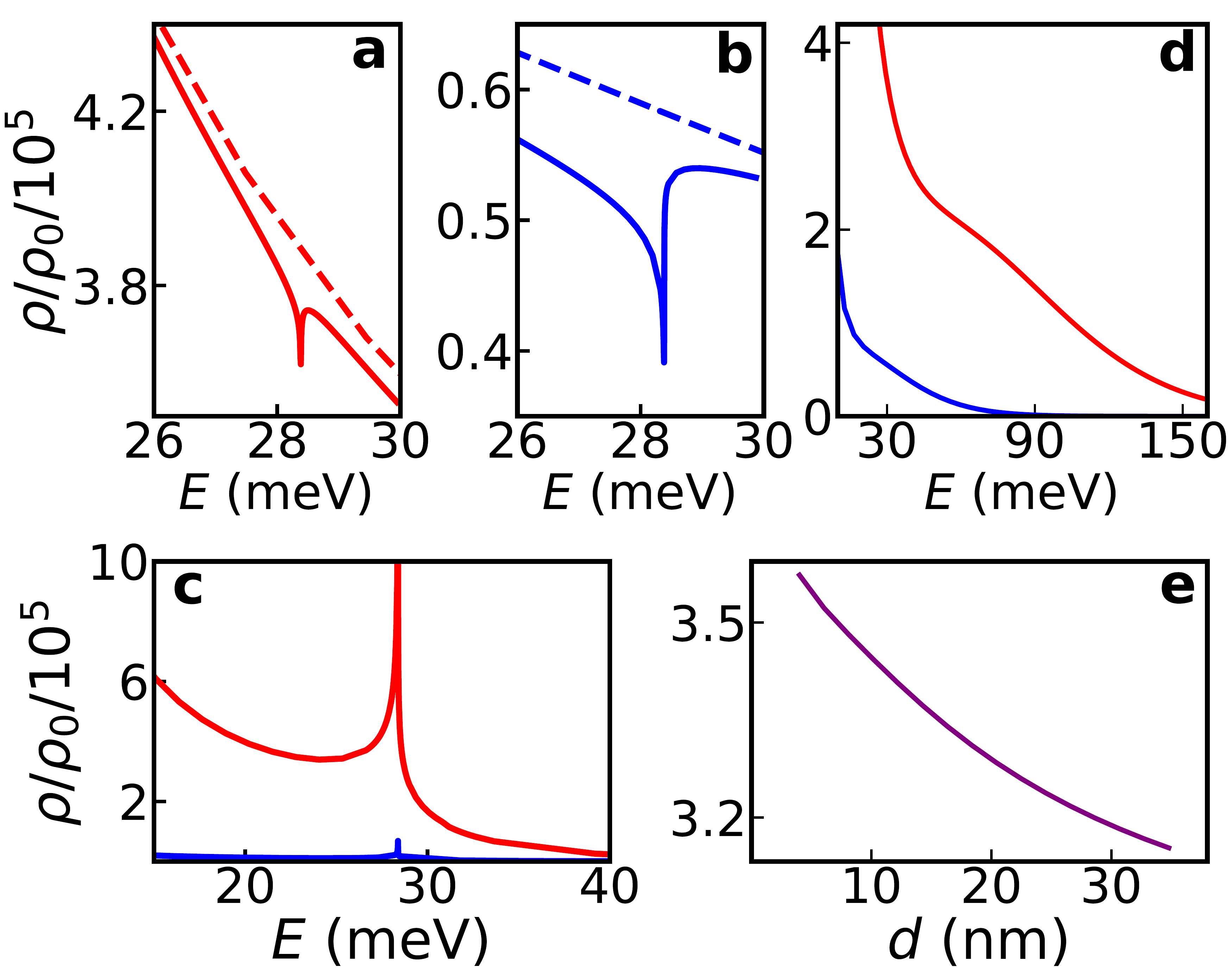}
    \caption{Purcell factor near a vacuum--hBN--graphene--hBN--superconductor heterostructure. In \textbf{a} and \textbf{b} the graphene Fermi energy has been set at $E_{\text{F}}^{\text{gr}} = \SI{0.25}{\eV}$; here, $T=\SI{1}{\kelvin}$ for the solid curves and $T=\SI{94}{\kelvin}$ (above $T_c$) for the dashed curves, and the graphene sheet is placed \SI{4}{\nm} above the superconductor's surface. We show results for two emitter--graphene distances: \SI{13}{\nm} (\textbf{a}) and \SI{36}{\nm} (\textbf{b}).  In \textbf{c} we show the case without graphene, at $T=\SI{1}{\kelvin}$. The red curve corresponds to an emitter--superconductor separation of \SI{17}{\nm} and the blue curve of \SI{40}{\nm}. In \textbf{d} we show results for the same distances as in \textbf{a} (red curve) and \textbf{b} (blue curve), but for $T=\SI{94}{\kelvin}$. In \textbf{e} we show how the Purcell factor depends on the graphene--superconductor distance $t$ at the energy of the Higgs mode, $\hbar\omega_{\text{\textsc{h}}} = 2\Delta_0 \approx \SI{28.32}{\meV}$. The other parameters are kept fixed: $E_{\text{F}}^{\text{gr}}=\SI{0.5}{\eV}$, $T=\SI{1}{\kelvin}$, and emitter--graphene distance of $d=\SI{13}{\nm}$. Here, graphene's conductivity has been modeled using the nonlocal RPA~\cite{Goncalves_SpringerTheses,Wunsch:2006}.
    }
    \label{fig:purcell_0-5eV}
\end{figure}

Figure~\ref{fig:purcell_0-5eV} shows the LDOS enhancement experienced by an emitter (or a nanosized tip) in the proximity of a superconductor; Figs.~\ref{fig:purcell_0-5eV}a--b,d--e refer to the case in the presence of graphene (located between the superconductor and the emitter), whereas Fig.~\ref{fig:purcell_0-5eV}c depicts a scenario where the graphene sheet is absent. The graphene sheet modifies the LDOS, affecting not only the absolute Purcell factor but also the peak/dip feature around the energy of the Higgs mode, $\hbar\omega_{\text{\textsc{h}}} = 2\Delta_0$. Such modification depends strongly on the emitter--graphene separation $d$ (Figs.~\ref{fig:purcell_0-5eV}a--b). Fig.~\ref{fig:purcell_0-5eV}d shows the LDOS enhancement for $T > T_c$ (\ie, above the superconductor's transition temperature) and thus the feature associated with the Higgs mode vanishes; all that remains is a relatively broad feature related to the excitation of graphene plasmons. 

\begin{figure}[t]
    \centering
    \includegraphics[width=\columnwidth]{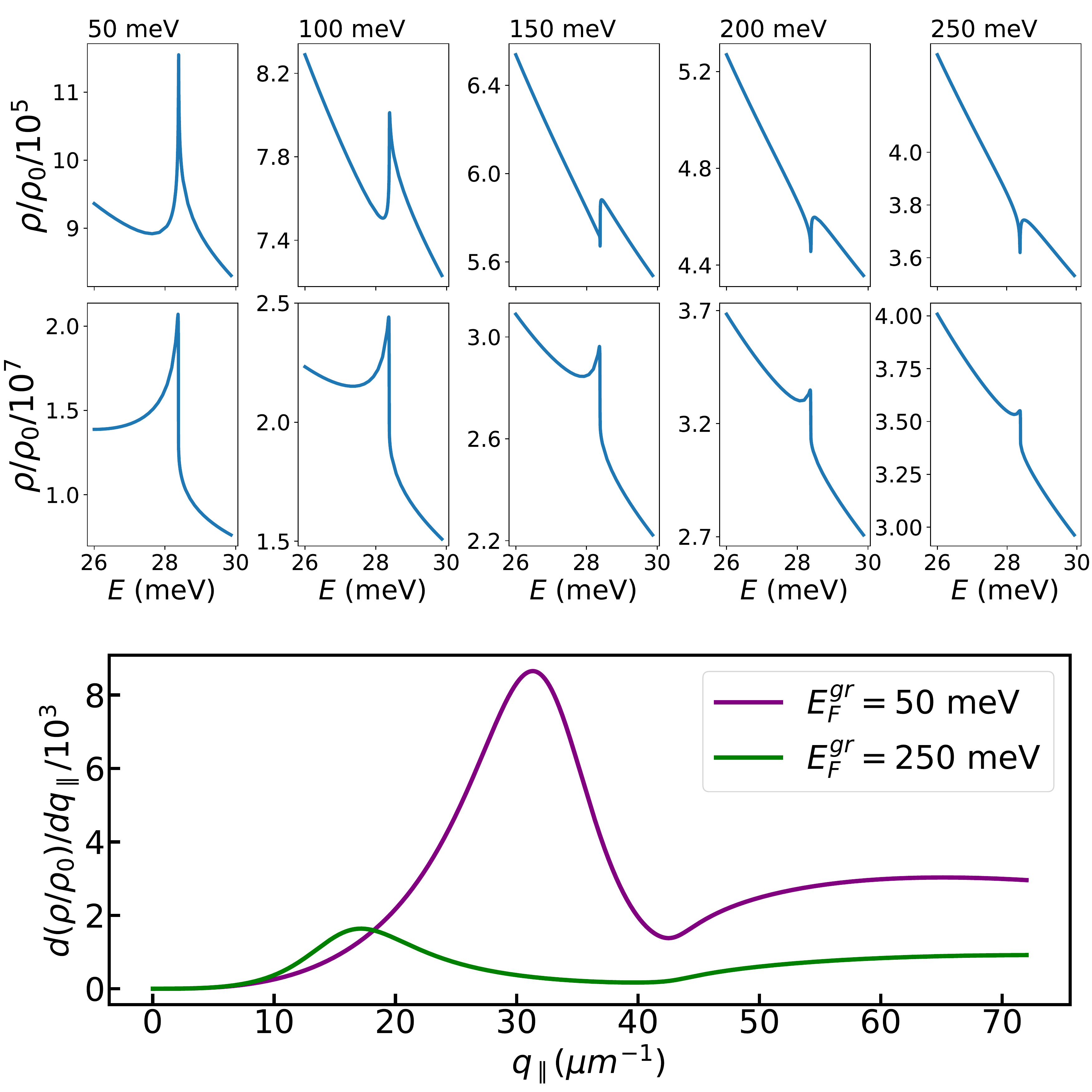}
    \caption{Purcell factor as a function of graphene's Fermi energy. Here we show the effect of changing graphene's
    Fermi energy (indicated at the top of each column) while keeping all other parameters fixed: $T=\SI{1}{\kelvin}$, emitter--graphene distance ($d=\SI{13}{\nm}$ for the top line and $d=\SI{2}{\nm}$ for the middle line), graphene--superconductor distance $t=\SI{4}{\nm}$. Here, graphene's conductivity has been modeled using the nonlocal RPA~\cite{Goncalves_SpringerTheses,Wunsch:2006}.
    For $d=\SI{13}{\nm}$, the dependence of the decay rate on the emitter's frequency changes quantitatively from low ($E_{\text{F}}^{\text{gr}}=\SI{50}{\meV}$) to high ($E_{\text{F}}^{\text{gr}}=\SI{250}{\meV}$) graphene doping. In the bottom panel we depict the $q_\parallel$-space differential LDOS given by the integration kernel of Eq.~(\ref{eq:rho}). The energy has been fixed at the value $\hbar\omega_{\mathrm{H}}$.}
    \label{fig:purcell_x_EFGr}
\end{figure}

Lastly, Fig.~\ref{fig:purcell_x_EFGr} depicts the LDOS enhancement for different values of graphene's Fermi energy (which can be tuned electrostatically), for two fixed emitter--graphene distances: $d=\SI{13}{\nm}$ (top row of panels) and $d=\SI{2}{\nm}$ (middle row of panels). 
For weakly doped graphene and the larger $d$ the sharp feature associated with the hybrid GPs-Higgs mode dominates the Purcell factor, being eventually overtaken by the broader background with increasing $E_{\text{F}}^{\text{gr}}$. To unveil the mechanisms underpinning the LDOS enhancement, we plot in the bottom row of Fig.~\ref{fig:purcell_x_EFGr} the $q_\parallel$-space differential LDOS enhancement [tantamount to the so-called $q_\parallel$-space power spectrum~\cite{FordWeber:1984}] , which is amounts to the integrand of Eq.~(\ref{eq:rho}). 
In the near-field (well-realized for the chosen setup and parameters), there are two contributions~\cite{FordWeber:1984,Novotny_book}: one from a resonant channel, corresponding to the excitation of the coupled Higgs--GP mode, and a broad, nonresonant contribution at larger $q_\parallel$ due to lossy channels (phenomenologically incorporated through the relaxation rates $\gamma,\gamma^{\text{gr}}$). Mathematically, the polariton (Higgs--GP mode) resonant contribution arises from the pole in $\Im r_p$, occurring at $q_\parallel \simeq \Re q_{\text{\textsc{GP}}}(\omega)$ [where $q_{\text{\textsc{GP}}}(\omega)$ is the wave vector of the Higgs--GP mode at frequency $\omega$ that satisfies the dispersion relation (cf.~Fig.~\ref{fig:G-SC_varying_EfGr-t})]. Consistent with this, the peak associated with the Higgs--GP polariton contribution to the $q_\parallel$-space differential LDOS occurs at a larger wave vector in the $E_{\text{F}}^{\text{gr}} = \SI{50}{\meV}$ case, since, for the same frequency, the Higgs--GP dispersion shifts toward larger wave vectors upon decreasing $E_{\text{F}}^{\text{gr}}$~\cite{Goncalves:2020b,Lundeberg:2017}. 
Ultimately, the amplitude of the resonant contribution depends on the specifics of the dispersion relation [i.e., $q_{\text{\textsc{GP}}}(\omega) = \Re q_{\text{\textsc{GP}}}(\omega) + \iu \Im q_{\text{\textsc{GP}}}(\omega)$], and is further weighted by the $q_\parallel^2 \exp\big(-2 q_\parallel z \big)$ factor that depends not only on the peak's location, $q_\parallel(\omega) \simeq \Re q_{\text{\textsc{GP}}}(\omega)$ [and whose width $\propto \Im q_{\text{\textsc{GP}}}(\omega)$], but also on the emitter's position $z=d-t'$ [\cf Eq.~(\ref{eq:rho})]. Finally, we stress that the relative contribution of each of the above-noted decay channels is strongly dependent on the emitter--graphene distance $d$ (with the nonresonant, lossy contribution eventually dominating at sufficiently small emitter--graphene separations---quenching)~\cite{FordWeber:1984,Novotny_book}.

\section*{Conclusion and Outlook}
\label{conclusion}

We have shown that signatures of a superconductor's Higgs mode can be detected by exploiting ultraconfined graphene plasmons supported by a graphene sheet placed in a superconductor's proximity. In particular, the presence of the Higgs mode for $T < T_c$ can be readily identified through an anticrossing feature that attests the coupling between graphene plasmons and the superconductor's Higgs mode. Further, we suggest that the excitation of the Higgs mode of superconductors could also be detected through the emergence of a peak or a dip in the near-field's Purcell factor, and whose shape (peak or dip) depends the coupling between the emitter and the continuum of the hybrid  GP--Higgs mode. This coupling is most efficient for small Fermi energies and short distances between the superconductor and the emitter.

Experimentally, the GP--Higgs interaction can be investigated using state-of-the-art cryogenic scanning near-field optical microscopy (SNOM)~\cite{Ni:2018}. Alternatively, more conventional spectroscopies relying on far-field optical techniques can also be explored by nanopatterning the graphene itself (e.g., into ribbon arrays) or its nearby materials (for example, the hBN or the superconductor). Examples of the latter---which have the benefit of preserving graphene from nanofabrication-induced defects---include the configurations studied in Refs.~\cite{Iranzo2018,Epstein:2020}, while the former approach can still be pursued using cutting-edge electron-beam lithography~\cite{Jessen:2019}. Another possibility is the use of highly-localized, local back-gate-free graphene doping modulation by placing a pristine graphene sheet on a substrate with patterned $\alpha$-$\text{RuCl}_3$~\cite{Rizzo:2020}.

Finally, there are a number of open questions that can spur from this work, \eg, if conductive thin films were added in direct electrical contact with the superconductor, then bound Andreev quasiparticle states inside the superconducting energy gap can form, being solutions to the Bogolubov--de Gennes equations~\cite{Sauls:2018}. Another enticing outlook is the prospect of using highly-confined GPs for investigating Josephson plasma waves in layered high-$T_c$ superconductors~\cite{Basov:2005,Dienst:2013,Laplace:2016}. The present formalism could be extended to the coupling of the above-noted types of modes (though this likely requires the use of more sophisticated models beyond the SCIB model employed here).

The work presented here sheds light on the fundamentals of collective excitations in novel architectures containing two-dimensional materials and superconductors and constitutes a proof-of-principle proposal, paving the way for prospective experimental investigations on the electrodynamics of superconductors using ultraconfined graphene plasmons.


\hfill

\vspace*{1cm}

\noindent
\textbf{Acknowledgments} %

\noindent
{\small 
N.\,M.\,R.\,P. acknowledges support from the European Commission through the project “Graphene-Driven Revolutions in ICT and Beyond” (Ref. No.~881603 -- Core 3), and the Portuguese Foundation for Science and Technology (FCT) in the framework of the Strategic Financing UID/FIS/04650/2019. N.\,M.\,R.\,P. also acknowledges COMPETE2020, PORTUGAL2020, FEDER, and the Portuguese Foundation for Science and Technology (FCT) through Project  No.~POCI-01-0145-FEDER-028114. 
N.\,A.\,M. is a VILLUM Investigator supported by VILLUM FONDEN (Grant No.~16498) and Independent Research Fund Denmark (Grant No.~7026-00117B).
The Center for Nano Optics is financially supported by the University
of Southern Denmark (SDU~2020 funding).
The Center for Nanostructured Graphene is sponsored by the Danish National Research Foundation (Project No.~DNRF103).
Work on hybrid heterostructures at Columbia was supported entirely by the Center on Precision-Assembled Quantum Materials, funded through the US National Science Foundation (NSF) Materials Research Science and Engineering Centers (award No.~DMR-2011738).  D.\,N.\,B. is Moore Investigator in Quantum Materials EPIQS~\#9455. D.\,N.\,B. is the Vannevar Bush Faculty Fellow ONR-VB: N00014-19-1-2630.
F.\,H.\,L.\,K. acknowledges financial support from the Government of Catalonia trough the SGR grant and from the Spanish Ministry of Economy and Competitiveness through the Severo Ochoa Programme for Centres of Excellence in R\&D (SEV-2015-0522), support by Fundaci\'{o} Cellex Barcelona, Generalitat de Catalunya through the CERCA program,  and the Mineco grants Plan Nacional (FIS2016-81044-P) and the Agency for Management of University and Research Grants (AGAUR) 2017 SGR 1656.  Furthermore, the research leading to these results has received funding from the European Union’s Horizon 2020 program under the Graphene Flagship grant agreements No.~785219 (Core~2) and No.~881603 (Core~3), and the Quantum Flagship grant No.~820378. This work was also supported by the ERC TOPONANOP under grant agreement No.~726001.
}

\vfill

\newpage
\bibliographystyle{apsrev4-1} 
\bibliography{references}

\begin{thebibliography}{67}%
\makeatletter
\providecommand \@ifxundefined [1]{%
 \@ifx{#1\undefined}
}%
\providecommand \@ifnum [1]{%
 \ifnum #1\expandafter \@firstoftwo
 \else \expandafter \@secondoftwo
 \fi
}%
\providecommand \@ifx [1]{%
 \ifx #1\expandafter \@firstoftwo
 \else \expandafter \@secondoftwo
 \fi
}%
\providecommand \natexlab [1]{#1}%
\providecommand \enquote  [1]{``#1''}%
\providecommand \bibnamefont  [1]{#1}%
\providecommand \bibfnamefont [1]{#1}%
\providecommand \citenamefont [1]{#1}%
\providecommand \href@noop [0]{\@secondoftwo}%
\providecommand \href [0]{\begingroup \@sanitize@url \@href}%
\providecommand \@href[1]{\@@startlink{#1}\@@href}%
\providecommand \@@href[1]{\endgroup#1\@@endlink}%
\providecommand \@sanitize@url [0]{\catcode `\\12\catcode `\$12\catcode
  `\&12\catcode `\#12\catcode `\^12\catcode `\_12\catcode `\%12\relax}%
\providecommand \@@startlink[1]{}%
\providecommand \@@endlink[0]{}%
\providecommand \url  [0]{\begingroup\@sanitize@url \@url }%
\providecommand \@url [1]{\endgroup\@href {#1}{\urlprefix }}%
\providecommand \urlprefix  [0]{URL }%
\providecommand \Eprint [0]{\href }%
\providecommand \doibase [0]{http://dx.doi.org/}%
\providecommand \selectlanguage [0]{\@gobble}%
\providecommand \bibinfo  [0]{\@secondoftwo}%
\providecommand \bibfield  [0]{\@secondoftwo}%
\providecommand \translation [1]{[#1]}%
\providecommand \BibitemOpen [0]{}%
\providecommand \bibitemStop [0]{}%
\providecommand \bibitemNoStop [0]{.\EOS\space}%
\providecommand \EOS [0]{\spacefactor3000\relax}%
\providecommand \BibitemShut  [1]{\csname bibitem#1\endcsname}%
\let\auto@bib@innerbib\@empty
\bibitem [{\citenamefont {Anderson}(1963)}]{Anderson1963}%
  \BibitemOpen
  \bibfield  {author} {\bibinfo {author} {\bibfnamefont {P.~W.}\ \bibnamefont
  {Anderson}},\ }\href {\doibase 10.1103/PhysRev.130.439} {\bibfield  {journal}
  {\bibinfo  {journal} {Phys. Rev.}\ }\textbf {\bibinfo {volume} {130}},\
  \bibinfo {pages} {439} (\bibinfo {year} {1963})}\BibitemShut {NoStop}%
\bibitem [{\citenamefont {Shimano}\ and\ \citenamefont
  {Tsuji}(2020)}]{Shimano2020}%
  \BibitemOpen
  \bibfield  {author} {\bibinfo {author} {\bibfnamefont {R.}~\bibnamefont
  {Shimano}}\ and\ \bibinfo {author} {\bibfnamefont {N.}~\bibnamefont
  {Tsuji}},\ }\href {\doibase 10.1146/annurev-conmatphys-031119-050813}
  {\bibfield  {journal} {\bibinfo  {journal} {Annu. Rev. Condens. Matter
  Phys.}\ }\textbf {\bibinfo {volume} {11}},\ \bibinfo {pages} {103} (\bibinfo
  {year} {2020})}\BibitemShut {NoStop}%
\bibitem [{\citenamefont {Pekker}\ and\ \citenamefont
  {Varma}(2015)}]{Pekker2015}%
  \BibitemOpen
  \bibfield  {author} {\bibinfo {author} {\bibfnamefont {D.}~\bibnamefont
  {Pekker}}\ and\ \bibinfo {author} {\bibfnamefont {C.}~\bibnamefont {Varma}},\
  }\href {\doibase 10.1146/annurev-conmatphys-031214-014350} {\bibfield
  {journal} {\bibinfo  {journal} {Annu. Rev. Condens. Matter Phys.}\ }\textbf
  {\bibinfo {volume} {6}},\ \bibinfo {pages} {269} (\bibinfo {year}
  {2015})}\BibitemShut {NoStop}%
\bibitem [{\citenamefont {Anderson}(2015)}]{Anderson2015}%
  \BibitemOpen
  \bibfield  {author} {\bibinfo {author} {\bibfnamefont {P.~W.}\ \bibnamefont
  {Anderson}},\ }\href {\doibase 10.1038/nphys3247} {\bibfield  {journal}
  {\bibinfo  {journal} {Nat. Physics}\ }\textbf {\bibinfo {volume} {11}},\
  \bibinfo {pages} {93} (\bibinfo {year} {2015})}\BibitemShut {NoStop}%
\bibitem [{\citenamefont {Higgs}(1964{\natexlab{a}})}]{Higgs1964a}%
  \BibitemOpen
  \bibfield  {author} {\bibinfo {author} {\bibfnamefont {P.~W.}\ \bibnamefont
  {Higgs}},\ }\href {\doibase https://doi.org/10.1016/0031-9163(64)91136-9}
  {\bibfield  {journal} {\bibinfo  {journal} {Phys. Lett.}\ }\textbf {\bibinfo
  {volume} {12}},\ \bibinfo {pages} {132 } (\bibinfo {year}
  {1964}{\natexlab{a}})}\BibitemShut {NoStop}%
\bibitem [{\citenamefont {Higgs}(1964{\natexlab{b}})}]{Higgs1964b}%
  \BibitemOpen
  \bibfield  {author} {\bibinfo {author} {\bibfnamefont {P.~W.}\ \bibnamefont
  {Higgs}},\ }\href {\doibase 10.1103/PhysRevLett.13.508} {\bibfield  {journal}
  {\bibinfo  {journal} {Phys. Rev. Lett.}\ }\textbf {\bibinfo {volume} {13}},\
  \bibinfo {pages} {508} (\bibinfo {year} {1964}{\natexlab{b}})}\BibitemShut
  {NoStop}%
\bibitem [{\citenamefont {Littlewood}\ and\ \citenamefont
  {Varma}(1982)}]{Littlewood1982}%
  \BibitemOpen
  \bibfield  {author} {\bibinfo {author} {\bibfnamefont {P.~B.}\ \bibnamefont
  {Littlewood}}\ and\ \bibinfo {author} {\bibfnamefont {C.~M.}\ \bibnamefont
  {Varma}},\ }\href {\doibase 10.1103/PhysRevB.26.4883} {\bibfield  {journal}
  {\bibinfo  {journal} {Phys. Rev. B}\ }\textbf {\bibinfo {volume} {26}},\
  \bibinfo {pages} {4883} (\bibinfo {year} {1982})}\BibitemShut {NoStop}%
\bibitem [{\citenamefont {Yang}\ and\ \citenamefont {Wu}(2019)}]{Yang2019}%
  \BibitemOpen
  \bibfield  {author} {\bibinfo {author} {\bibfnamefont {F.}~\bibnamefont
  {Yang}}\ and\ \bibinfo {author} {\bibfnamefont {M.~W.}\ \bibnamefont {Wu}},\
  }\href {\doibase 10.1103/PhysRevB.100.104513} {\bibfield  {journal} {\bibinfo
   {journal} {Phys. Rev. B}\ }\textbf {\bibinfo {volume} {100}},\ \bibinfo
  {pages} {104513} (\bibinfo {year} {2019})}\BibitemShut {NoStop}%
\bibitem [{Com()}]{CommentHiggsDetection}%
  \BibitemOpen
  \href@noop {} {}\bibinfo {note} {It has been recently suggested, however,
  that the observed oscillations could be interpreted as resulting from
  excitation of the NG mode
  instead~\cite{Yang2019,Yang2018,Tsuchiya2018,Yu2017a,Yu2017b}. Additionally,
  it has also been pointed out that the Higgs mode may be observed in
  disordered superconductors~\cite{Sherman2015}, as long as one chooses to
  measure the appropriate response function~\cite{Podolsky2011}.}\BibitemShut
  {Stop}%
\bibitem [{\citenamefont {Matsunaga}\ and\ \citenamefont
  {Shimano}(2012)}]{Matsunaga2012}%
  \BibitemOpen
  \bibfield  {author} {\bibinfo {author} {\bibfnamefont {R.}~\bibnamefont
  {Matsunaga}}\ and\ \bibinfo {author} {\bibfnamefont {R.}~\bibnamefont
  {Shimano}},\ }\href {\doibase 10.1103/PhysRevLett.109.187002} {\bibfield
  {journal} {\bibinfo  {journal} {Phys. Rev. Lett.}\ }\textbf {\bibinfo
  {volume} {109}},\ \bibinfo {pages} {187002} (\bibinfo {year}
  {2012})}\BibitemShut {NoStop}%
\bibitem [{\citenamefont {Matsunaga}\ \emph {et~al.}(2013)\citenamefont
  {Matsunaga}, \citenamefont {Hamada}, \citenamefont {Makise}, \citenamefont
  {Uzawa}, \citenamefont {Terai}, \citenamefont {Wang},\ and\ \citenamefont
  {Shimano}}]{Matsunaga2013}%
  \BibitemOpen
  \bibfield  {author} {\bibinfo {author} {\bibfnamefont {R.}~\bibnamefont
  {Matsunaga}}, \bibinfo {author} {\bibfnamefont {Y.~I.}\ \bibnamefont
  {Hamada}}, \bibinfo {author} {\bibfnamefont {K.}~\bibnamefont {Makise}},
  \bibinfo {author} {\bibfnamefont {Y.}~\bibnamefont {Uzawa}}, \bibinfo
  {author} {\bibfnamefont {H.}~\bibnamefont {Terai}}, \bibinfo {author}
  {\bibfnamefont {Z.}~\bibnamefont {Wang}}, \ and\ \bibinfo {author}
  {\bibfnamefont {R.}~\bibnamefont {Shimano}},\ }\href {\doibase
  10.1103/PhysRevLett.111.057002} {\bibfield  {journal} {\bibinfo  {journal}
  {Phys. Rev. Lett.}\ }\textbf {\bibinfo {volume} {111}},\ \bibinfo {pages}
  {057002} (\bibinfo {year} {2013})}\BibitemShut {NoStop}%
\bibitem [{\citenamefont {Matsunaga}\ \emph {et~al.}(2014)\citenamefont
  {Matsunaga}, \citenamefont {Tsuji}, \citenamefont {Fujita}, \citenamefont
  {Sugioka}, \citenamefont {Makise}, \citenamefont {Uzawa}, \citenamefont
  {Terai}, \citenamefont {Wang}, \citenamefont {Aoki},\ and\ \citenamefont
  {Shimano}}]{Matsunaga2014}%
  \BibitemOpen
  \bibfield  {author} {\bibinfo {author} {\bibfnamefont {R.}~\bibnamefont
  {Matsunaga}}, \bibinfo {author} {\bibfnamefont {N.}~\bibnamefont {Tsuji}},
  \bibinfo {author} {\bibfnamefont {H.}~\bibnamefont {Fujita}}, \bibinfo
  {author} {\bibfnamefont {A.}~\bibnamefont {Sugioka}}, \bibinfo {author}
  {\bibfnamefont {K.}~\bibnamefont {Makise}}, \bibinfo {author} {\bibfnamefont
  {Y.}~\bibnamefont {Uzawa}}, \bibinfo {author} {\bibfnamefont
  {H.}~\bibnamefont {Terai}}, \bibinfo {author} {\bibfnamefont
  {Z.}~\bibnamefont {Wang}}, \bibinfo {author} {\bibfnamefont {H.}~\bibnamefont
  {Aoki}}, \ and\ \bibinfo {author} {\bibfnamefont {R.}~\bibnamefont
  {Shimano}},\ }\href {\doibase 10.1126/science.1254697} {\bibfield  {journal}
  {\bibinfo  {journal} {Science}\ }\textbf {\bibinfo {volume} {345}},\ \bibinfo
  {pages} {1145} (\bibinfo {year} {2014})}\BibitemShut {NoStop}%
\bibitem [{\citenamefont {Matsunaga}\ \emph {et~al.}(2017)\citenamefont
  {Matsunaga}, \citenamefont {Tsuji}, \citenamefont {Makise}, \citenamefont
  {Terai}, \citenamefont {Aoki},\ and\ \citenamefont
  {Shimano}}]{Matsunaga2017}%
  \BibitemOpen
  \bibfield  {author} {\bibinfo {author} {\bibfnamefont {R.}~\bibnamefont
  {Matsunaga}}, \bibinfo {author} {\bibfnamefont {N.}~\bibnamefont {Tsuji}},
  \bibinfo {author} {\bibfnamefont {K.}~\bibnamefont {Makise}}, \bibinfo
  {author} {\bibfnamefont {H.}~\bibnamefont {Terai}}, \bibinfo {author}
  {\bibfnamefont {H.}~\bibnamefont {Aoki}}, \ and\ \bibinfo {author}
  {\bibfnamefont {R.}~\bibnamefont {Shimano}},\ }\href {\doibase
  10.1103/PhysRevB.96.020505} {\bibfield  {journal} {\bibinfo  {journal} {Phys.
  Rev. B}\ }\textbf {\bibinfo {volume} {96}},\ \bibinfo {pages} {020505(R)}
  (\bibinfo {year} {2017})}\BibitemShut {NoStop}%
\bibitem [{\citenamefont {Katsumi}\ \emph {et~al.}(2018)\citenamefont
  {Katsumi}, \citenamefont {Tsuji}, \citenamefont {Hamada}, \citenamefont
  {Matsunaga}, \citenamefont {Schneeloch}, \citenamefont {Zhong}, \citenamefont
  {Gu}, \citenamefont {Aoki}, \citenamefont {Gallais},\ and\ \citenamefont
  {Shimano}}]{Katsumi2018}%
  \BibitemOpen
  \bibfield  {author} {\bibinfo {author} {\bibfnamefont {K.}~\bibnamefont
  {Katsumi}}, \bibinfo {author} {\bibfnamefont {N.}~\bibnamefont {Tsuji}},
  \bibinfo {author} {\bibfnamefont {Y.~I.}\ \bibnamefont {Hamada}}, \bibinfo
  {author} {\bibfnamefont {R.}~\bibnamefont {Matsunaga}}, \bibinfo {author}
  {\bibfnamefont {J.}~\bibnamefont {Schneeloch}}, \bibinfo {author}
  {\bibfnamefont {R.~D.}\ \bibnamefont {Zhong}}, \bibinfo {author}
  {\bibfnamefont {G.~D.}\ \bibnamefont {Gu}}, \bibinfo {author} {\bibfnamefont
  {H.}~\bibnamefont {Aoki}}, \bibinfo {author} {\bibfnamefont {Y.}~\bibnamefont
  {Gallais}}, \ and\ \bibinfo {author} {\bibfnamefont {R.}~\bibnamefont
  {Shimano}},\ }\href {\doibase 10.1103/PhysRevLett.120.117001} {\bibfield
  {journal} {\bibinfo  {journal} {Phys. Rev. Lett.}\ }\textbf {\bibinfo
  {volume} {120}},\ \bibinfo {pages} {117001} (\bibinfo {year}
  {2018})}\BibitemShut {NoStop}%
\bibitem [{\citenamefont {Sun}\ \emph {et~al.}(2020)\citenamefont {Sun},
  \citenamefont {Fogler}, \citenamefont {Basov},\ and\ \citenamefont
  {Millis}}]{sun2020}%
  \BibitemOpen
  \bibfield  {author} {\bibinfo {author} {\bibfnamefont {Z.}~\bibnamefont
  {Sun}}, \bibinfo {author} {\bibfnamefont {M.~M.}\ \bibnamefont {Fogler}},
  \bibinfo {author} {\bibfnamefont {D.~N.}\ \bibnamefont {Basov}}, \ and\
  \bibinfo {author} {\bibfnamefont {A.~J.}\ \bibnamefont {Millis}},\ }\href
  {\doibase 10.1103/PhysRevResearch.2.023413} {\bibfield  {journal} {\bibinfo
  {journal} {Phys. Rev. Research}\ }\textbf {\bibinfo {volume} {2}},\ \bibinfo
  {pages} {023413} (\bibinfo {year} {2020})}\BibitemShut {NoStop}%
\bibitem [{\citenamefont {Rickayzen}(1965)}]{rickayzen1965theory}%
  \BibitemOpen
  \bibfield  {author} {\bibinfo {author} {\bibfnamefont {G.}~\bibnamefont
  {Rickayzen}},\ }\href@noop {} {\emph {\bibinfo {title} {Theory of
  superconductivity}}}\ (\bibinfo  {publisher} {Interscience Publishers},\
  \bibinfo {address} {New York},\ \bibinfo {year} {1965})\BibitemShut {NoStop}%
\bibitem [{\citenamefont {Gon\c{c}alves}\ and\ \citenamefont
  {Peres}(2016)}]{GoncalvesPeresBook}%
  \BibitemOpen
  \bibfield  {author} {\bibinfo {author} {\bibfnamefont {P.~A.~D.}\
  \bibnamefont {Gon\c{c}alves}}\ and\ \bibinfo {author} {\bibfnamefont
  {N.~M.~R.}\ \bibnamefont {Peres}},\ }\href {\doibase 10.1142/9948} {\emph
  {\bibinfo {title} {An Introduction to Graphene Plasmonics}}}\ (\bibinfo
  {publisher} {World Scientific},\ \bibinfo {address} {Singapore},\ \bibinfo
  {year} {2016})\BibitemShut {NoStop}%
\bibitem [{\citenamefont {Gon\c{c}alves}(2020)}]{Goncalves_SpringerTheses}%
  \BibitemOpen
  \bibfield  {author} {\bibinfo {author} {\bibfnamefont {P.~A.~D.}\
  \bibnamefont {Gon\c{c}alves}},\ }\href {\doibase 10.1007/978-3-030-38291-9}
  {\emph {\bibinfo {title} {Plasmonics and Light--Matter Interactions in
  Two-Dimensional Materials and in Metal Nanostructures: Classical and Quantum
  Considerations}}}\ (\bibinfo  {publisher} {Springer Nature},\ \bibinfo {year}
  {2020})\BibitemShut {NoStop}%
\bibitem [{\citenamefont {Lundeberg}\ \emph {et~al.}(2017)\citenamefont
  {Lundeberg}, \citenamefont {Gao}, \citenamefont {Asgari}, \citenamefont
  {Tan}, \citenamefont {Van~Duppen}, \citenamefont {Autore}, \citenamefont
  {Alonso-Gonz{\'a}lez}, \citenamefont {Woessner}, \citenamefont {Watanabe},
  \citenamefont {Taniguchi}, \citenamefont {Hillenbrand}, \citenamefont {Hone},
  \citenamefont {Polini},\ and\ \citenamefont {Koppens}}]{Lundeberg:2017}%
  \BibitemOpen
  \bibfield  {author} {\bibinfo {author} {\bibfnamefont {M.~B.}\ \bibnamefont
  {Lundeberg}}, \bibinfo {author} {\bibfnamefont {Y.}~\bibnamefont {Gao}},
  \bibinfo {author} {\bibfnamefont {R.}~\bibnamefont {Asgari}}, \bibinfo
  {author} {\bibfnamefont {C.}~\bibnamefont {Tan}}, \bibinfo {author}
  {\bibfnamefont {B.}~\bibnamefont {Van~Duppen}}, \bibinfo {author}
  {\bibfnamefont {M.}~\bibnamefont {Autore}}, \bibinfo {author} {\bibfnamefont
  {P.}~\bibnamefont {Alonso-Gonz{\'a}lez}}, \bibinfo {author} {\bibfnamefont
  {A.}~\bibnamefont {Woessner}}, \bibinfo {author} {\bibfnamefont
  {K.}~\bibnamefont {Watanabe}}, \bibinfo {author} {\bibfnamefont
  {T.}~\bibnamefont {Taniguchi}}, \bibinfo {author} {\bibfnamefont
  {R.}~\bibnamefont {Hillenbrand}}, \bibinfo {author} {\bibfnamefont
  {J.}~\bibnamefont {Hone}}, \bibinfo {author} {\bibfnamefont {M.}~\bibnamefont
  {Polini}}, \ and\ \bibinfo {author} {\bibfnamefont {F.~H.~L.}\ \bibnamefont
  {Koppens}},\ }\href {\doibase 10.1126/science.aan2735} {\bibfield  {journal}
  {\bibinfo  {journal} {Science}\ }\textbf {\bibinfo {volume} {357}},\ \bibinfo
  {pages} {187} (\bibinfo {year} {2017})}\BibitemShut {NoStop}%
\bibitem [{\citenamefont {Alcaraz~Iranzo}\ \emph {et~al.}(2018)\citenamefont
  {Alcaraz~Iranzo}, \citenamefont {Nanot}, \citenamefont {Dias}, \citenamefont
  {Epstein}, \citenamefont {Peng}, \citenamefont {Efetov}, \citenamefont
  {Lundeberg}, \citenamefont {Parret}, \citenamefont {Osmond}, \citenamefont
  {Hong}, \citenamefont {Kong}, \citenamefont {Englund}, \citenamefont
  {Peres},\ and\ \citenamefont {Koppens}}]{Iranzo2018}%
  \BibitemOpen
  \bibfield  {author} {\bibinfo {author} {\bibfnamefont {D.}~\bibnamefont
  {Alcaraz~Iranzo}}, \bibinfo {author} {\bibfnamefont {S.}~\bibnamefont
  {Nanot}}, \bibinfo {author} {\bibfnamefont {E.~J.~C.}\ \bibnamefont {Dias}},
  \bibinfo {author} {\bibfnamefont {I.}~\bibnamefont {Epstein}}, \bibinfo
  {author} {\bibfnamefont {C.}~\bibnamefont {Peng}}, \bibinfo {author}
  {\bibfnamefont {D.~K.}\ \bibnamefont {Efetov}}, \bibinfo {author}
  {\bibfnamefont {M.~B.}\ \bibnamefont {Lundeberg}}, \bibinfo {author}
  {\bibfnamefont {R.}~\bibnamefont {Parret}}, \bibinfo {author} {\bibfnamefont
  {J.}~\bibnamefont {Osmond}}, \bibinfo {author} {\bibfnamefont {J.-Y.}\
  \bibnamefont {Hong}}, \bibinfo {author} {\bibfnamefont {J.}~\bibnamefont
  {Kong}}, \bibinfo {author} {\bibfnamefont {D.~R.}\ \bibnamefont {Englund}},
  \bibinfo {author} {\bibfnamefont {N.~M.~R.}\ \bibnamefont {Peres}}, \ and\
  \bibinfo {author} {\bibfnamefont {F.~H.~L.}\ \bibnamefont {Koppens}},\ }\href
  {\doibase 10.1126/science.aar8438} {\bibfield  {journal} {\bibinfo  {journal}
  {Science}\ }\textbf {\bibinfo {volume} {360}},\ \bibinfo {pages} {291}
  (\bibinfo {year} {2018})}\BibitemShut {NoStop}%
\bibitem [{\citenamefont {Dias}\ \emph {et~al.}(2018)\citenamefont {Dias},
  \citenamefont {Iranzo}, \citenamefont {Gon\ifmmode~\mbox{\c{c}}\else
  \c{c}\fi{}alves}, \citenamefont {Hajati}, \citenamefont {Bludov},
  \citenamefont {Jauho}, \citenamefont {Mortensen}, \citenamefont {Koppens},\
  and\ \citenamefont {Peres}}]{Dias2018}%
  \BibitemOpen
  \bibfield  {author} {\bibinfo {author} {\bibfnamefont {E.~J.~C.}\
  \bibnamefont {Dias}}, \bibinfo {author} {\bibfnamefont {D.~A.}\ \bibnamefont
  {Iranzo}}, \bibinfo {author} {\bibfnamefont {P.~A.~D.}\ \bibnamefont
  {Gon\ifmmode~\mbox{\c{c}}\else \c{c}\fi{}alves}}, \bibinfo {author}
  {\bibfnamefont {Y.}~\bibnamefont {Hajati}}, \bibinfo {author} {\bibfnamefont
  {Y.~V.}\ \bibnamefont {Bludov}}, \bibinfo {author} {\bibfnamefont {A.-P.}\
  \bibnamefont {Jauho}}, \bibinfo {author} {\bibfnamefont {N.~A.}\ \bibnamefont
  {Mortensen}}, \bibinfo {author} {\bibfnamefont {F.~H.~L.}\ \bibnamefont
  {Koppens}}, \ and\ \bibinfo {author} {\bibfnamefont {N.~M.~R.}\ \bibnamefont
  {Peres}},\ }\href {\doibase 10.1103/PhysRevB.97.245405} {\bibfield  {journal}
  {\bibinfo  {journal} {Phys. Rev. B}\ }\textbf {\bibinfo {volume} {97}},\
  \bibinfo {pages} {245405} (\bibinfo {year} {2018})}\BibitemShut {NoStop}%
\bibitem [{\citenamefont {Gon\c{c}alves}\ \emph
  {et~al.}(2020{\natexlab{a}})\citenamefont {Gon\c{c}alves}, \citenamefont
  {Christensen}, \citenamefont {Peres}, \citenamefont {Jauho}, \citenamefont
  {Epstein}, \citenamefont {Koppens}, \citenamefont {Solja\v{c}i\'{c}},\ and\
  \citenamefont {Mortensen}}]{Goncalves:2020a}%
  \BibitemOpen
  \bibfield  {author} {\bibinfo {author} {\bibfnamefont {P.~A.~D.}\
  \bibnamefont {Gon\c{c}alves}}, \bibinfo {author} {\bibfnamefont
  {T.}~\bibnamefont {Christensen}}, \bibinfo {author} {\bibfnamefont
  {N.~M.~R.}\ \bibnamefont {Peres}}, \bibinfo {author} {\bibfnamefont {A.-P.}\
  \bibnamefont {Jauho}}, \bibinfo {author} {\bibfnamefont {I.}~\bibnamefont
  {Epstein}}, \bibinfo {author} {\bibfnamefont {F.~H.~L.}\ \bibnamefont
  {Koppens}}, \bibinfo {author} {\bibfnamefont {M.}~\bibnamefont
  {Solja\v{c}i\'{c}}}, \ and\ \bibinfo {author} {\bibfnamefont {N.~A.}\
  \bibnamefont {Mortensen}},\ }\href {https://arxiv.org/abs/2008.07613}
  {\bibfield  {journal} {\bibinfo  {journal} {arXiv:2008.07613}\ } (\bibinfo
  {year} {2020}{\natexlab{a}})}\BibitemShut {NoStop}%
\bibitem [{\citenamefont {Gon\c{c}alves}\ \emph
  {et~al.}(2020{\natexlab{b}})\citenamefont {Gon\c{c}alves}, \citenamefont
  {Stenger}, \citenamefont {Cox}, \citenamefont {Mortensen},\ and\
  \citenamefont {Xiao}}]{Goncalves:2020b}%
  \BibitemOpen
  \bibfield  {author} {\bibinfo {author} {\bibfnamefont {P.~A.~D.}\
  \bibnamefont {Gon\c{c}alves}}, \bibinfo {author} {\bibfnamefont
  {N.}~\bibnamefont {Stenger}}, \bibinfo {author} {\bibfnamefont {J.~D.}\
  \bibnamefont {Cox}}, \bibinfo {author} {\bibfnamefont {N.~A.}\ \bibnamefont
  {Mortensen}}, \ and\ \bibinfo {author} {\bibfnamefont {S.}~\bibnamefont
  {Xiao}},\ }\href {\doibase 10.1002/adom.201901473} {\bibfield  {journal}
  {\bibinfo  {journal} {Adv. Opt. Mater.}\ }\textbf {\bibinfo {volume} {8}},\
  \bibinfo {pages} {1901473} (\bibinfo {year}
  {2020}{\natexlab{b}})}\BibitemShut {NoStop}%
\bibitem [{\citenamefont {Basov}\ and\ \citenamefont
  {Timusk}(2005)}]{Basov:2005}%
  \BibitemOpen
  \bibfield  {author} {\bibinfo {author} {\bibfnamefont {D.~N.}\ \bibnamefont
  {Basov}}\ and\ \bibinfo {author} {\bibfnamefont {T.}~\bibnamefont {Timusk}},\
  }\href {\doibase 10.1103/RevModPhys.77.721} {\bibfield  {journal} {\bibinfo
  {journal} {Rev. Mod. Phys.}\ }\textbf {\bibinfo {volume} {77}},\ \bibinfo
  {pages} {721} (\bibinfo {year} {2005})}\BibitemShut {NoStop}%
\bibitem [{\citenamefont {Basov}\ \emph {et~al.}(2011)\citenamefont {Basov},
  \citenamefont {Averitt}, \citenamefont {van~der Marel}, \citenamefont
  {Dressel},\ and\ \citenamefont {Haule}}]{Basov:2011}%
  \BibitemOpen
  \bibfield  {author} {\bibinfo {author} {\bibfnamefont {D.~N.}\ \bibnamefont
  {Basov}}, \bibinfo {author} {\bibfnamefont {R.~D.}\ \bibnamefont {Averitt}},
  \bibinfo {author} {\bibfnamefont {D.}~\bibnamefont {van~der Marel}}, \bibinfo
  {author} {\bibfnamefont {M.}~\bibnamefont {Dressel}}, \ and\ \bibinfo
  {author} {\bibfnamefont {K.}~\bibnamefont {Haule}},\ }\href {\doibase
  10.1103/RevModPhys.83.471} {\bibfield  {journal} {\bibinfo  {journal} {Rev.
  Mod. Phys.}\ }\textbf {\bibinfo {volume} {83}},\ \bibinfo {pages} {471}
  (\bibinfo {year} {2011})}\BibitemShut {NoStop}%
\bibitem [{\citenamefont {Bouscher}\ \emph {et~al.}(2017)\citenamefont
  {Bouscher}, \citenamefont {Panna},\ and\ \citenamefont
  {Hayat}}]{Bouscher2017}%
  \BibitemOpen
  \bibfield  {author} {\bibinfo {author} {\bibfnamefont {S.}~\bibnamefont
  {Bouscher}}, \bibinfo {author} {\bibfnamefont {D.}~\bibnamefont {Panna}}, \
  and\ \bibinfo {author} {\bibfnamefont {A.}~\bibnamefont {Hayat}},\ }\href
  {\doibase 10.1088/2040-8986/aa8888} {\bibfield  {journal} {\bibinfo
  {journal} {J. Opt.}\ }\textbf {\bibinfo {volume} {19}},\ \bibinfo {pages}
  {103003} (\bibinfo {year} {2017})}\BibitemShut {NoStop}%
\bibitem [{\citenamefont {Epstein}\ \emph {et~al.}(2020)\citenamefont
  {Epstein}, \citenamefont {Alcaraz}, \citenamefont {Huang}, \citenamefont
  {Pusapati}, \citenamefont {Hugonin}, \citenamefont {Kumar}, \citenamefont
  {Deputy}, \citenamefont {Khodkov}, \citenamefont {Rappoport}, \citenamefont
  {Peres}, \citenamefont {Smith},\ and\ \citenamefont
  {Koppens}}]{Epstein:2020}%
  \BibitemOpen
  \bibfield  {author} {\bibinfo {author} {\bibfnamefont {I.}~\bibnamefont
  {Epstein}}, \bibinfo {author} {\bibfnamefont {D.}~\bibnamefont {Alcaraz}},
  \bibinfo {author} {\bibfnamefont {Z.}~\bibnamefont {Huang}}, \bibinfo
  {author} {\bibfnamefont {V.-V.}\ \bibnamefont {Pusapati}}, \bibinfo {author}
  {\bibfnamefont {J.-P.}\ \bibnamefont {Hugonin}}, \bibinfo {author}
  {\bibfnamefont {A.}~\bibnamefont {Kumar}}, \bibinfo {author} {\bibfnamefont
  {X.}~\bibnamefont {Deputy}}, \bibinfo {author} {\bibfnamefont
  {T.}~\bibnamefont {Khodkov}}, \bibinfo {author} {\bibfnamefont {T.~G.}\
  \bibnamefont {Rappoport}}, \bibinfo {author} {\bibfnamefont {N.~M.~R.}\
  \bibnamefont {Peres}}, \bibinfo {author} {\bibfnamefont {D.~R.}\ \bibnamefont
  {Smith}}, \ and\ \bibinfo {author} {\bibfnamefont {F.~H.~L.}\ \bibnamefont
  {Koppens}},\ }\href {\doibase 10.1126/science.abb1570} {\bibfield  {journal}
  {\bibinfo  {journal} {Science}\ }\textbf {\bibinfo {volume} {368}},\ \bibinfo
  {pages} {1219} (\bibinfo {year} {2020})}\BibitemShut {NoStop}%
\bibitem [{\citenamefont {Purcell}(1946)}]{Purcell:1946}%
  \BibitemOpen
  \bibfield  {author} {\bibinfo {author} {\bibfnamefont {E.~M.}\ \bibnamefont
  {Purcell}},\ }\href {\doibase 10.1103/PhysRev.69.674.2} {\bibfield  {journal}
  {\bibinfo  {journal} {Phys. Rev.}\ }\textbf {\bibinfo {volume} {69}},\
  \bibinfo {pages} {681} (\bibinfo {year} {1946})}\BibitemShut {NoStop}%
\bibitem [{\citenamefont {Novotny}\ and\ \citenamefont
  {Hecht}(2012)}]{Novotny_book}%
  \BibitemOpen
  \bibfield  {author} {\bibinfo {author} {\bibfnamefont {L.}~\bibnamefont
  {Novotny}}\ and\ \bibinfo {author} {\bibfnamefont {B.}~\bibnamefont
  {Hecht}},\ }\href {\doibase 10.1017/CBO9780511794193} {\emph {\bibinfo
  {title} {Principles of Nano-Optics}}},\ \bibinfo {edition} {2nd}\ ed.\
  (\bibinfo  {publisher} {Cambridge University Press},\ \bibinfo {year}
  {2012})\BibitemShut {NoStop}%
\bibitem [{\citenamefont {Bardeen}\ \emph {et~al.}(1957)\citenamefont
  {Bardeen}, \citenamefont {Cooper},\ and\ \citenamefont
  {Schrieffer}}]{Bardeen:1957}%
  \BibitemOpen
  \bibfield  {author} {\bibinfo {author} {\bibfnamefont {J.}~\bibnamefont
  {Bardeen}}, \bibinfo {author} {\bibfnamefont {L.~N.}\ \bibnamefont {Cooper}},
  \ and\ \bibinfo {author} {\bibfnamefont {J.~R.}\ \bibnamefont {Schrieffer}},\
  }\href {\doibase 10.1103/PhysRev.108.1175} {\bibfield  {journal} {\bibinfo
  {journal} {Phys. Rev.}\ }\textbf {\bibinfo {volume} {108}},\ \bibinfo {pages}
  {1175} (\bibinfo {year} {1957})}\BibitemShut {NoStop}%
\bibitem [{\citenamefont {Schrieffer}(1999)}]{Schrieffer_SCbook}%
  \BibitemOpen
  \bibfield  {author} {\bibinfo {author} {\bibfnamefont {J.~R.}\ \bibnamefont
  {Schrieffer}},\ }\href {\doibase 10.1201/9780429495700} {\emph {\bibinfo
  {title} {Theory of Superconductivity}}},\ Advanced Books Classics\ (\bibinfo
  {publisher} {CRC Press},\ \bibinfo {year} {1999})\BibitemShut {NoStop}%
\bibitem [{sig()}]{sigmaHomogeneous}%
  \BibitemOpen
  \href@noop {} {}\bibinfo {note} {In translationally invariant, homogeneous
  media, the linear optical conductivity tensor satisfies
  $\tensorGrk{\sigma}(\mathbf{q},\omega) =
  \tensorGrk{\sigma}(-\mathbf{q},\omega)$. Consequently, under such assumption,
  and for $q \ll k_{\text{F}}$, the lowest-order nonlocal correction to the
  optical conductivity is in second order in $q$.}\BibitemShut {Stop}%
\bibitem [{\citenamefont {Keller}(1990)}]{Keller1990}%
  \BibitemOpen
  \bibfield  {author} {\bibinfo {author} {\bibfnamefont {O.}~\bibnamefont
  {Keller}},\ }\href {\doibase 10.1364/JOSAB.7.002229} {\bibfield  {journal}
  {\bibinfo  {journal} {J. Opt. Soc. Am. B}\ }\textbf {\bibinfo {volume} {7}},\
  \bibinfo {pages} {2229} (\bibinfo {year} {1990})}\BibitemShut {NoStop}%
\bibitem [{\citenamefont {Keller}\ and\ \citenamefont
  {Liu}(1991)}]{Keller:1991}%
  \BibitemOpen
  \bibfield  {author} {\bibinfo {author} {\bibfnamefont {O.}~\bibnamefont
  {Keller}}\ and\ \bibinfo {author} {\bibfnamefont {A.}~\bibnamefont {Liu}},\
  }\href {\doibase 10.1016/0030-4018(91)90256-D} {\bibfield  {journal}
  {\bibinfo  {journal} {Opt. Commun.}\ }\textbf {\bibinfo {volume} {80}},\
  \bibinfo {pages} {229} (\bibinfo {year} {1991})}\BibitemShut {NoStop}%
\bibitem [{\citenamefont {Ford}\ and\ \citenamefont
  {Weber}(1984)}]{FordWeber:1984}%
  \BibitemOpen
  \bibfield  {author} {\bibinfo {author} {\bibfnamefont {G.~W.}\ \bibnamefont
  {Ford}}\ and\ \bibinfo {author} {\bibfnamefont {W.~H.}\ \bibnamefont
  {Weber}},\ }\href {\doibase https://doi.org/10.1016/0370-1573(84)90098-X}
  {\bibfield  {journal} {\bibinfo  {journal} {Phys. Rep.}\ }\textbf {\bibinfo
  {volume} {113}},\ \bibinfo {pages} {195} (\bibinfo {year}
  {1984})}\BibitemShut {NoStop}%
\bibitem [{\citenamefont {Keller}\ and\ \citenamefont
  {Pedersen}(1989)}]{Keller1989}%
  \BibitemOpen
  \bibfield  {author} {\bibinfo {author} {\bibfnamefont {O.}~\bibnamefont
  {Keller}}\ and\ \bibinfo {author} {\bibfnamefont {J.~H.}\ \bibnamefont
  {Pedersen}},\ }\href {\doibase 10.1117/12.950359} {\bibfield  {journal}
  {\bibinfo  {journal} {Proc. SPIE}\ }\textbf {\bibinfo {volume} {1029}},\
  \bibinfo {pages} {18} (\bibinfo {year} {1989})}\BibitemShut {NoStop}%
\bibitem [{\citenamefont {Wesche}(2015)}]{PhysPropHighTcBook}%
  \BibitemOpen
  \bibfield  {author} {\bibinfo {author} {\bibfnamefont {R.}~\bibnamefont
  {Wesche}},\ }\href {\doibase 10.1002/9781118696644} {\emph {\bibinfo {title}
  {Physical Properties of High‐Temperature Superconductors}}}\ (\bibinfo
  {publisher} {John Wiley \& Sons, Ltd},\ \bibinfo {year} {2015})\BibitemShut
  {NoStop}%
\bibitem [{\citenamefont {Jackson}(1998)}]{Jackson}%
  \BibitemOpen
  \bibfield  {author} {\bibinfo {author} {\bibfnamefont {J.~D.}\ \bibnamefont
  {Jackson}},\ }\href@noop {} {\emph {\bibinfo {title} {Classical
  Electrodynamics}}},\ \bibinfo {edition} {3rd}\ ed.\ (\bibinfo  {publisher}
  {Wiley},\ \bibinfo {address} {New York},\ \bibinfo {year} {1998})\BibitemShut
  {NoStop}%
\bibitem [{\citenamefont {Ni}\ \emph {et~al.}(2018)\citenamefont {Ni},
  \citenamefont {McLeod}, \citenamefont {Sun}, \citenamefont {Wang},
  \citenamefont {Xiong}, \citenamefont {Post}, \citenamefont {Sunku},
  \citenamefont {Jiang}, \citenamefont {Hone}, \citenamefont {Dean},
  \citenamefont {Fogler},\ and\ \citenamefont {Basov}}]{Ni:2018}%
  \BibitemOpen
  \bibfield  {author} {\bibinfo {author} {\bibfnamefont {G.~X.}\ \bibnamefont
  {Ni}}, \bibinfo {author} {\bibfnamefont {A.~S.}\ \bibnamefont {McLeod}},
  \bibinfo {author} {\bibfnamefont {Z.}~\bibnamefont {Sun}}, \bibinfo {author}
  {\bibfnamefont {L.}~\bibnamefont {Wang}}, \bibinfo {author} {\bibfnamefont
  {L.}~\bibnamefont {Xiong}}, \bibinfo {author} {\bibfnamefont {K.~W.}\
  \bibnamefont {Post}}, \bibinfo {author} {\bibfnamefont {S.~S.}\ \bibnamefont
  {Sunku}}, \bibinfo {author} {\bibfnamefont {B.-Y.}\ \bibnamefont {Jiang}},
  \bibinfo {author} {\bibfnamefont {J.}~\bibnamefont {Hone}}, \bibinfo {author}
  {\bibfnamefont {C.~R.}\ \bibnamefont {Dean}}, \bibinfo {author}
  {\bibfnamefont {M.~M.}\ \bibnamefont {Fogler}}, \ and\ \bibinfo {author}
  {\bibfnamefont {D.~N.}\ \bibnamefont {Basov}},\ }\href {\doibase
  10.1038/s41586-018-0136-9} {\bibfield  {journal} {\bibinfo  {journal}
  {Nature}\ }\textbf {\bibinfo {volume} {557}},\ \bibinfo {pages} {530}
  (\bibinfo {year} {2018})}\BibitemShut {NoStop}%
\bibitem [{\citenamefont {Stockman}\ \emph {et~al.}(2018)\citenamefont
  {Stockman}, \citenamefont {Kneipp}, \citenamefont {Bozhevolnyi},
  \citenamefont {Saha}, \citenamefont {Dutta}, \citenamefont {Ndukaife},
  \citenamefont {Kinsey}, \citenamefont {Reddy}, \citenamefont {Guler},
  \citenamefont {Shalaev}, \citenamefont {Boltasseva}, \citenamefont
  {Gholipour}, \citenamefont {Krishnamoorthy}, \citenamefont {MacDonald},
  \citenamefont {Soci}, \citenamefont {Zheludev}, \citenamefont {Savinov},
  \citenamefont {Singh}, \citenamefont {Gro{\ss}}, \citenamefont {Lienau},
  \citenamefont {Vadai}, \citenamefont {Solomon}, \citenamefont {Barton},
  \citenamefont {Lawrence}, \citenamefont {Dionne}, \citenamefont {Boriskina},
  \citenamefont {Esteban}, \citenamefont {Aizpurua}, \citenamefont {Zhang},
  \citenamefont {Yang}, \citenamefont {Wang}, \citenamefont {Wang},
  \citenamefont {Odom}, \citenamefont {Accanto}, \citenamefont {de~Roque},
  \citenamefont {Hancu}, \citenamefont {Piatkowski}, \citenamefont {van
  Hulst},\ and\ \citenamefont {Kling}}]{Stockman:2018}%
  \BibitemOpen
  \bibfield  {author} {\bibinfo {author} {\bibfnamefont {M.~I.}\ \bibnamefont
  {Stockman}}, \bibinfo {author} {\bibfnamefont {K.}~\bibnamefont {Kneipp}},
  \bibinfo {author} {\bibfnamefont {S.~I.}\ \bibnamefont {Bozhevolnyi}},
  \bibinfo {author} {\bibfnamefont {S.}~\bibnamefont {Saha}}, \bibinfo {author}
  {\bibfnamefont {A.}~\bibnamefont {Dutta}}, \bibinfo {author} {\bibfnamefont
  {J.}~\bibnamefont {Ndukaife}}, \bibinfo {author} {\bibfnamefont
  {N.}~\bibnamefont {Kinsey}}, \bibinfo {author} {\bibfnamefont
  {H.}~\bibnamefont {Reddy}}, \bibinfo {author} {\bibfnamefont
  {U.}~\bibnamefont {Guler}}, \bibinfo {author} {\bibfnamefont {V.~M.}\
  \bibnamefont {Shalaev}}, \bibinfo {author} {\bibfnamefont {A.}~\bibnamefont
  {Boltasseva}}, \bibinfo {author} {\bibfnamefont {B.}~\bibnamefont
  {Gholipour}}, \bibinfo {author} {\bibfnamefont {H.~N.~S.}\ \bibnamefont
  {Krishnamoorthy}}, \bibinfo {author} {\bibfnamefont {K.~F.}\ \bibnamefont
  {MacDonald}}, \bibinfo {author} {\bibfnamefont {C.}~\bibnamefont {Soci}},
  \bibinfo {author} {\bibfnamefont {N.~I.}\ \bibnamefont {Zheludev}}, \bibinfo
  {author} {\bibfnamefont {V.}~\bibnamefont {Savinov}}, \bibinfo {author}
  {\bibfnamefont {R.}~\bibnamefont {Singh}}, \bibinfo {author} {\bibfnamefont
  {P.}~\bibnamefont {Gro{\ss}}}, \bibinfo {author} {\bibfnamefont
  {C.}~\bibnamefont {Lienau}}, \bibinfo {author} {\bibfnamefont
  {M.}~\bibnamefont {Vadai}}, \bibinfo {author} {\bibfnamefont {M.~L.}\
  \bibnamefont {Solomon}}, \bibinfo {author} {\bibfnamefont {D.~R.}\
  \bibnamefont {Barton}}, \bibinfo {author} {\bibfnamefont {M.}~\bibnamefont
  {Lawrence}}, \bibinfo {author} {\bibfnamefont {J.~A.}\ \bibnamefont
  {Dionne}}, \bibinfo {author} {\bibfnamefont {S.~V.}\ \bibnamefont
  {Boriskina}}, \bibinfo {author} {\bibfnamefont {R.}~\bibnamefont {Esteban}},
  \bibinfo {author} {\bibfnamefont {J.}~\bibnamefont {Aizpurua}}, \bibinfo
  {author} {\bibfnamefont {X.}~\bibnamefont {Zhang}}, \bibinfo {author}
  {\bibfnamefont {S.}~\bibnamefont {Yang}}, \bibinfo {author} {\bibfnamefont
  {D.}~\bibnamefont {Wang}}, \bibinfo {author} {\bibfnamefont {W.}~\bibnamefont
  {Wang}}, \bibinfo {author} {\bibfnamefont {T.~W.}\ \bibnamefont {Odom}},
  \bibinfo {author} {\bibfnamefont {N.}~\bibnamefont {Accanto}}, \bibinfo
  {author} {\bibfnamefont {P.~M.}\ \bibnamefont {de~Roque}}, \bibinfo {author}
  {\bibfnamefont {I.~M.}\ \bibnamefont {Hancu}}, \bibinfo {author}
  {\bibfnamefont {L.}~\bibnamefont {Piatkowski}}, \bibinfo {author}
  {\bibfnamefont {N.~F.}\ \bibnamefont {van Hulst}}, \ and\ \bibinfo {author}
  {\bibfnamefont {M.~F.}\ \bibnamefont {Kling}},\ }\href {\doibase
  10.1088/2040-8986/aaa114} {\bibfield  {journal} {\bibinfo  {journal} {J.
  Opt.}\ }\textbf {\bibinfo {volume} {20}},\ \bibinfo {pages} {043001}
  (\bibinfo {year} {2018})}\BibitemShut {NoStop}%
\bibitem [{\citenamefont {Stinson}\ \emph {et~al.}(2014)\citenamefont
  {Stinson}, \citenamefont {Wu}, \citenamefont {Jiang}, \citenamefont {Fei},
  \citenamefont {Rodin}, \citenamefont {Chapler}, \citenamefont {McLeod},
  \citenamefont {Castro~Neto}, \citenamefont {Lee}, \citenamefont {Fogler},\
  and\ \citenamefont {Basov}}]{Stinson:2014}%
  \BibitemOpen
  \bibfield  {author} {\bibinfo {author} {\bibfnamefont {H.~T.}\ \bibnamefont
  {Stinson}}, \bibinfo {author} {\bibfnamefont {J.~S.}\ \bibnamefont {Wu}},
  \bibinfo {author} {\bibfnamefont {B.~Y.}\ \bibnamefont {Jiang}}, \bibinfo
  {author} {\bibfnamefont {Z.}~\bibnamefont {Fei}}, \bibinfo {author}
  {\bibfnamefont {A.~S.}\ \bibnamefont {Rodin}}, \bibinfo {author}
  {\bibfnamefont {B.~C.}\ \bibnamefont {Chapler}}, \bibinfo {author}
  {\bibfnamefont {A.~S.}\ \bibnamefont {McLeod}}, \bibinfo {author}
  {\bibfnamefont {A.}~\bibnamefont {Castro~Neto}}, \bibinfo {author}
  {\bibfnamefont {Y.~S.}\ \bibnamefont {Lee}}, \bibinfo {author} {\bibfnamefont
  {M.~M.}\ \bibnamefont {Fogler}}, \ and\ \bibinfo {author} {\bibfnamefont
  {D.~N.}\ \bibnamefont {Basov}},\ }\href {\doibase 10.1103/PhysRevB.90.014502}
  {\bibfield  {journal} {\bibinfo  {journal} {Phys. Rev. B}\ }\textbf {\bibinfo
  {volume} {90}},\ \bibinfo {pages} {014502} (\bibinfo {year}
  {2014})}\BibitemShut {NoStop}%
\bibitem [{\citenamefont {Basov}\ \emph {et~al.}(2016)\citenamefont {Basov},
  \citenamefont {Fogler},\ and\ \citenamefont {Garc{\'\i}a~de
  Abajo}}]{Basov:2016}%
  \BibitemOpen
  \bibfield  {author} {\bibinfo {author} {\bibfnamefont {D.~N.}\ \bibnamefont
  {Basov}}, \bibinfo {author} {\bibfnamefont {M.~M.}\ \bibnamefont {Fogler}}, \
  and\ \bibinfo {author} {\bibfnamefont {F.~J.}\ \bibnamefont {Garc{\'\i}a~de
  Abajo}},\ }\href {\doibase 10.1126/science.aag1992} {\bibfield  {journal}
  {\bibinfo  {journal} {Science}\ }\textbf {\bibinfo {volume} {354}},\ \bibinfo
  {pages} {aag1992} (\bibinfo {year} {2016})}\BibitemShut {NoStop}%
\bibitem [{\citenamefont {Chen}\ \emph {et~al.}(2012)\citenamefont {Chen},
  \citenamefont {Badioli}, \citenamefont {Alonso-González}, \citenamefont
  {Thongrattanasiri}, \citenamefont {Huth}, \citenamefont {Osmond},
  \citenamefont {Spasenović}, \citenamefont {Centeno}, \citenamefont
  {Pesquera}, \citenamefont {Godignon}, \citenamefont {Elorza}, \citenamefont
  {Camara}, \citenamefont {{García de Abajo}}, \citenamefont {Hillenbrand},\
  and\ \citenamefont {Koppens}}]{Chen:2012}%
  \BibitemOpen
  \bibfield  {author} {\bibinfo {author} {\bibfnamefont {J.}~\bibnamefont
  {Chen}}, \bibinfo {author} {\bibfnamefont {M.}~\bibnamefont {Badioli}},
  \bibinfo {author} {\bibfnamefont {P.}~\bibnamefont {Alonso-González}},
  \bibinfo {author} {\bibfnamefont {S.}~\bibnamefont {Thongrattanasiri}},
  \bibinfo {author} {\bibfnamefont {F.}~\bibnamefont {Huth}}, \bibinfo {author}
  {\bibfnamefont {J.}~\bibnamefont {Osmond}}, \bibinfo {author} {\bibfnamefont
  {M.}~\bibnamefont {Spasenović}}, \bibinfo {author} {\bibfnamefont
  {A.}~\bibnamefont {Centeno}}, \bibinfo {author} {\bibfnamefont
  {A.}~\bibnamefont {Pesquera}}, \bibinfo {author} {\bibfnamefont
  {P.}~\bibnamefont {Godignon}}, \bibinfo {author} {\bibfnamefont {A.~Z.}\
  \bibnamefont {Elorza}}, \bibinfo {author} {\bibfnamefont {N.}~\bibnamefont
  {Camara}}, \bibinfo {author} {\bibfnamefont {F.~J.}\ \bibnamefont {{García
  de Abajo}}}, \bibinfo {author} {\bibfnamefont {R.}~\bibnamefont
  {Hillenbrand}}, \ and\ \bibinfo {author} {\bibfnamefont {F.~H.~L.}\
  \bibnamefont {Koppens}},\ }\href {\doibase 10.1038/nature11254} {\bibfield
  {journal} {\bibinfo  {journal} {Nature}\ }\textbf {\bibinfo {volume} {487}},\
  \bibinfo {pages} {77} (\bibinfo {year} {2012})}\BibitemShut {NoStop}%
\bibitem [{\citenamefont {Fei}\ \emph {et~al.}(2012)\citenamefont {Fei},
  \citenamefont {Rodin}, \citenamefont {Andreev}, \citenamefont {Bao},
  \citenamefont {McLeod}, \citenamefont {Wagner}, \citenamefont {Zhang},
  \citenamefont {Zhao}, \citenamefont {Thiemens}, \citenamefont {Dominguez},
  \citenamefont {Fogler}, \citenamefont {Neto}, \citenamefont {Lau},
  \citenamefont {Keilmann},\ and\ \citenamefont {Basov}}]{Fei2012}%
  \BibitemOpen
  \bibfield  {author} {\bibinfo {author} {\bibfnamefont {Z.}~\bibnamefont
  {Fei}}, \bibinfo {author} {\bibfnamefont {A.~S.}\ \bibnamefont {Rodin}},
  \bibinfo {author} {\bibfnamefont {G.~O.}\ \bibnamefont {Andreev}}, \bibinfo
  {author} {\bibfnamefont {W.}~\bibnamefont {Bao}}, \bibinfo {author}
  {\bibfnamefont {A.~S.}\ \bibnamefont {McLeod}}, \bibinfo {author}
  {\bibfnamefont {M.}~\bibnamefont {Wagner}}, \bibinfo {author} {\bibfnamefont
  {L.~M.}\ \bibnamefont {Zhang}}, \bibinfo {author} {\bibfnamefont
  {Z.}~\bibnamefont {Zhao}}, \bibinfo {author} {\bibfnamefont {M.}~\bibnamefont
  {Thiemens}}, \bibinfo {author} {\bibfnamefont {G.}~\bibnamefont {Dominguez}},
  \bibinfo {author} {\bibfnamefont {M.~M.}\ \bibnamefont {Fogler}}, \bibinfo
  {author} {\bibfnamefont {A.~H.~C.}\ \bibnamefont {Neto}}, \bibinfo {author}
  {\bibfnamefont {C.~N.}\ \bibnamefont {Lau}}, \bibinfo {author} {\bibfnamefont
  {F.}~\bibnamefont {Keilmann}}, \ and\ \bibinfo {author} {\bibfnamefont
  {D.~N.}\ \bibnamefont {Basov}},\ }\href {\doibase 10.1038/nature11253}
  {\bibfield  {journal} {\bibinfo  {journal} {Nature}\ }\textbf {\bibinfo
  {volume} {487}},\ \bibinfo {pages} {82} (\bibinfo {year} {2012})}\BibitemShut
  {NoStop}%
\bibitem [{\citenamefont {Woessner}\ \emph {et~al.}(2015)\citenamefont
  {Woessner}, \citenamefont {Lundeberg}, \citenamefont {Gao}, \citenamefont
  {Principi}, \citenamefont {Alonso-Gonz{\'a}lez}, \citenamefont {Carrega},
  \citenamefont {Watanabe}, \citenamefont {Taniguchi}, \citenamefont {Vignale},
  \citenamefont {Polini}, \citenamefont {Hone}, \citenamefont {Hillenbrand},\
  and\ \citenamefont {Koppens}}]{Woessner2015}%
  \BibitemOpen
  \bibfield  {author} {\bibinfo {author} {\bibfnamefont {A.}~\bibnamefont
  {Woessner}}, \bibinfo {author} {\bibfnamefont {M.~B.}\ \bibnamefont
  {Lundeberg}}, \bibinfo {author} {\bibfnamefont {Y.}~\bibnamefont {Gao}},
  \bibinfo {author} {\bibfnamefont {A.}~\bibnamefont {Principi}}, \bibinfo
  {author} {\bibfnamefont {P.}~\bibnamefont {Alonso-Gonz{\'a}lez}}, \bibinfo
  {author} {\bibfnamefont {M.}~\bibnamefont {Carrega}}, \bibinfo {author}
  {\bibfnamefont {K.}~\bibnamefont {Watanabe}}, \bibinfo {author}
  {\bibfnamefont {T.}~\bibnamefont {Taniguchi}}, \bibinfo {author}
  {\bibfnamefont {G.}~\bibnamefont {Vignale}}, \bibinfo {author} {\bibfnamefont
  {M.}~\bibnamefont {Polini}}, \bibinfo {author} {\bibfnamefont
  {J.}~\bibnamefont {Hone}}, \bibinfo {author} {\bibfnamefont {R.}~\bibnamefont
  {Hillenbrand}}, \ and\ \bibinfo {author} {\bibfnamefont {F.~H.~L.}\
  \bibnamefont {Koppens}},\ }\href {\doibase 10.1038/nmat4169} {\bibfield
  {journal} {\bibinfo  {journal} {Nat. Mater.}\ }\textbf {\bibinfo {volume}
  {14}},\ \bibinfo {pages} {421} (\bibinfo {year} {2015})}\BibitemShut
  {NoStop}%
\bibitem [{\citenamefont {Cai}\ \emph {et~al.}(2007)\citenamefont {Cai},
  \citenamefont {Zhang}, \citenamefont {Zeng}, \citenamefont {Cheng},\ and\
  \citenamefont {Xu}}]{Cai:2007}%
  \BibitemOpen
  \bibfield  {author} {\bibinfo {author} {\bibfnamefont {Y.}~\bibnamefont
  {Cai}}, \bibinfo {author} {\bibfnamefont {L.}~\bibnamefont {Zhang}}, \bibinfo
  {author} {\bibfnamefont {Q.}~\bibnamefont {Zeng}}, \bibinfo {author}
  {\bibfnamefont {L.}~\bibnamefont {Cheng}}, \ and\ \bibinfo {author}
  {\bibfnamefont {Y.}~\bibnamefont {Xu}},\ }\href {\doibase
  10.1016/j.ssc.2006.10.040} {\bibfield  {journal} {\bibinfo  {journal} {Solid
  State Commun.}\ }\textbf {\bibinfo {volume} {141}},\ \bibinfo {pages} {262}
  (\bibinfo {year} {2007})}\BibitemShut {NoStop}%
\bibitem [{\citenamefont {Koppens}\ \emph {et~al.}(2011)\citenamefont
  {Koppens}, \citenamefont {Chang},\ and\ \citenamefont {Garc{\'i}a~de
  Abajo}}]{Koppens2011}%
  \BibitemOpen
  \bibfield  {author} {\bibinfo {author} {\bibfnamefont {F.~H.~L.}\
  \bibnamefont {Koppens}}, \bibinfo {author} {\bibfnamefont {D.~E.}\
  \bibnamefont {Chang}}, \ and\ \bibinfo {author} {\bibfnamefont {F.~J.}\
  \bibnamefont {Garc{\'i}a~de Abajo}},\ }\href {\doibase 10.1021/nl201771h}
  {\bibfield  {journal} {\bibinfo  {journal} {Nano Lett.}\ }\textbf {\bibinfo
  {volume} {11}},\ \bibinfo {pages} {3370} (\bibinfo {year}
  {2011})}\BibitemShut {NoStop}%
\bibitem [{\citenamefont {Sch{\"a}dler}\ \emph {et~al.}(2019)\citenamefont
  {Sch{\"a}dler}, \citenamefont {Ciancico}, \citenamefont {Pazzagli},
  \citenamefont {Lombardi}, \citenamefont {Bachtold}, \citenamefont
  {Toninelli}, \citenamefont {Reserbat-Plantey},\ and\ \citenamefont
  {Koppens}}]{Schadler2019}%
  \BibitemOpen
  \bibfield  {author} {\bibinfo {author} {\bibfnamefont {K.~G.}\ \bibnamefont
  {Sch{\"a}dler}}, \bibinfo {author} {\bibfnamefont {C.}~\bibnamefont
  {Ciancico}}, \bibinfo {author} {\bibfnamefont {S.}~\bibnamefont {Pazzagli}},
  \bibinfo {author} {\bibfnamefont {P.}~\bibnamefont {Lombardi}}, \bibinfo
  {author} {\bibfnamefont {A.}~\bibnamefont {Bachtold}}, \bibinfo {author}
  {\bibfnamefont {C.}~\bibnamefont {Toninelli}}, \bibinfo {author}
  {\bibfnamefont {A.}~\bibnamefont {Reserbat-Plantey}}, \ and\ \bibinfo
  {author} {\bibfnamefont {F.~H.~L.}\ \bibnamefont {Koppens}},\ }\href
  {\doibase 10.1021/acs.nanolett.9b00916} {\bibfield  {journal} {\bibinfo
  {journal} {Nano Lett.}\ }\textbf {\bibinfo {volume} {19}},\ \bibinfo {pages}
  {3789} (\bibinfo {year} {2019})}\BibitemShut {NoStop}%
\bibitem [{\citenamefont {Kurman}\ \emph {et~al.}(2018)\citenamefont {Kurman},
  \citenamefont {Rivera}, \citenamefont {Christensen}, \citenamefont {Tsesses},
  \citenamefont {Orenstein}, \citenamefont {Solja{\v{c}}i{\'{c}}},
  \citenamefont {Joannopoulos},\ and\ \citenamefont {Kaminer}}]{Kurman2018}%
  \BibitemOpen
  \bibfield  {author} {\bibinfo {author} {\bibfnamefont {Y.}~\bibnamefont
  {Kurman}}, \bibinfo {author} {\bibfnamefont {N.}~\bibnamefont {Rivera}},
  \bibinfo {author} {\bibfnamefont {T.}~\bibnamefont {Christensen}}, \bibinfo
  {author} {\bibfnamefont {S.}~\bibnamefont {Tsesses}}, \bibinfo {author}
  {\bibfnamefont {M.}~\bibnamefont {Orenstein}}, \bibinfo {author}
  {\bibfnamefont {M.}~\bibnamefont {Solja{\v{c}}i{\'{c}}}}, \bibinfo {author}
  {\bibfnamefont {J.~D.}\ \bibnamefont {Joannopoulos}}, \ and\ \bibinfo
  {author} {\bibfnamefont {I.}~\bibnamefont {Kaminer}},\ }\href {\doibase
  10.1038/s41566-018-0176-6} {\bibfield  {journal} {\bibinfo  {journal} {Nat.
  Photon.}\ }\textbf {\bibinfo {volume} {12}},\ \bibinfo {pages} {423}
  (\bibinfo {year} {2018})}\BibitemShut {NoStop}%
\bibitem [{\citenamefont {Scarafagio}\ \emph {et~al.}(2019)\citenamefont
  {Scarafagio}, \citenamefont {Tallaire}, \citenamefont {Tielrooij},
  \citenamefont {Cano}, \citenamefont {Grishin}, \citenamefont {Chavanne},
  \citenamefont {Koppens}, \citenamefont {Ringuedé}, \citenamefont {Cassir},
  \citenamefont {Serrano}, \citenamefont {Goldner},\ and\ \citenamefont
  {Ferrier}}]{Scarafagio:2019}%
  \BibitemOpen
  \bibfield  {author} {\bibinfo {author} {\bibfnamefont {M.}~\bibnamefont
  {Scarafagio}}, \bibinfo {author} {\bibfnamefont {A.}~\bibnamefont
  {Tallaire}}, \bibinfo {author} {\bibfnamefont {K.-J.}\ \bibnamefont
  {Tielrooij}}, \bibinfo {author} {\bibfnamefont {D.}~\bibnamefont {Cano}},
  \bibinfo {author} {\bibfnamefont {A.}~\bibnamefont {Grishin}}, \bibinfo
  {author} {\bibfnamefont {M.-H.}\ \bibnamefont {Chavanne}}, \bibinfo {author}
  {\bibfnamefont {F.~H.~L.}\ \bibnamefont {Koppens}}, \bibinfo {author}
  {\bibfnamefont {A.}~\bibnamefont {Ringuedé}}, \bibinfo {author}
  {\bibfnamefont {M.}~\bibnamefont {Cassir}}, \bibinfo {author} {\bibfnamefont
  {D.}~\bibnamefont {Serrano}}, \bibinfo {author} {\bibfnamefont
  {P.}~\bibnamefont {Goldner}}, \ and\ \bibinfo {author} {\bibfnamefont
  {A.}~\bibnamefont {Ferrier}},\ }\href {\doibase 10.1021/acs.jpcc.9b02597}
  {\bibfield  {journal} {\bibinfo  {journal} {J. Phys. Chem. C.}\ }\textbf
  {\bibinfo {volume} {123}},\ \bibinfo {pages} {13354} (\bibinfo {year}
  {2019})}\BibitemShut {NoStop}%
\bibitem [{\citenamefont {Cano}\ \emph {et~al.}(2020)\citenamefont {Cano},
  \citenamefont {Ferrier}, \citenamefont {Soundarapandian}, \citenamefont
  {Reserbat-Plantey}, \citenamefont {Scarafagio}, \citenamefont {Tallaire},
  \citenamefont {Seyeux}, \citenamefont {Marcus}, \citenamefont
  {de~Riedmatten}, \citenamefont {Goldner}, \citenamefont {Koppens},\ and\
  \citenamefont {Tielrooij}}]{Cano:2020}%
  \BibitemOpen
  \bibfield  {author} {\bibinfo {author} {\bibfnamefont {D.}~\bibnamefont
  {Cano}}, \bibinfo {author} {\bibfnamefont {A.}~\bibnamefont {Ferrier}},
  \bibinfo {author} {\bibfnamefont {K.}~\bibnamefont {Soundarapandian}},
  \bibinfo {author} {\bibfnamefont {A.}~\bibnamefont {Reserbat-Plantey}},
  \bibinfo {author} {\bibfnamefont {M.}~\bibnamefont {Scarafagio}}, \bibinfo
  {author} {\bibfnamefont {A.}~\bibnamefont {Tallaire}}, \bibinfo {author}
  {\bibfnamefont {A.}~\bibnamefont {Seyeux}}, \bibinfo {author} {\bibfnamefont
  {P.}~\bibnamefont {Marcus}}, \bibinfo {author} {\bibfnamefont
  {H.}~\bibnamefont {de~Riedmatten}}, \bibinfo {author} {\bibfnamefont
  {P.}~\bibnamefont {Goldner}}, \bibinfo {author} {\bibfnamefont {F.~H.~L.}\
  \bibnamefont {Koppens}}, \ and\ \bibinfo {author} {\bibfnamefont {K.-J.}\
  \bibnamefont {Tielrooij}},\ }\href {\doibase 10.1038/s41467-020-17899-7}
  {\bibfield  {journal} {\bibinfo  {journal} {Nat. Commun.}\ }\textbf {\bibinfo
  {volume} {11}},\ \bibinfo {pages} {4094} (\bibinfo {year}
  {2020})}\BibitemShut {NoStop}%
\bibitem [{\citenamefont {Zibik}\ \emph {et~al.}(2009)\citenamefont {Zibik},
  \citenamefont {Grange}, \citenamefont {Carpenter}, \citenamefont {Porter},
  \citenamefont {Ferreira}, \citenamefont {Bastard}, \citenamefont {Stehr},
  \citenamefont {Winnerl}, \citenamefont {Helm}, \citenamefont {Liu},
  \citenamefont {Skolnick},\ and\ \citenamefont {Wilson}}]{Zibik2009}%
  \BibitemOpen
  \bibfield  {author} {\bibinfo {author} {\bibfnamefont {E.~A.}\ \bibnamefont
  {Zibik}}, \bibinfo {author} {\bibfnamefont {T.}~\bibnamefont {Grange}},
  \bibinfo {author} {\bibfnamefont {B.~A.}\ \bibnamefont {Carpenter}}, \bibinfo
  {author} {\bibfnamefont {N.~E.}\ \bibnamefont {Porter}}, \bibinfo {author}
  {\bibfnamefont {R.}~\bibnamefont {Ferreira}}, \bibinfo {author}
  {\bibfnamefont {G.}~\bibnamefont {Bastard}}, \bibinfo {author} {\bibfnamefont
  {D.}~\bibnamefont {Stehr}}, \bibinfo {author} {\bibfnamefont
  {S.}~\bibnamefont {Winnerl}}, \bibinfo {author} {\bibfnamefont
  {M.}~\bibnamefont {Helm}}, \bibinfo {author} {\bibfnamefont {H.~Y.}\
  \bibnamefont {Liu}}, \bibinfo {author} {\bibfnamefont {M.~S.}\ \bibnamefont
  {Skolnick}}, \ and\ \bibinfo {author} {\bibfnamefont {L.~R.}\ \bibnamefont
  {Wilson}},\ }\href {\doibase 10.1038/nmat2511} {\bibfield  {journal}
  {\bibinfo  {journal} {Nat. Mater.}\ }\textbf {\bibinfo {volume} {8}},\
  \bibinfo {pages} {803} (\bibinfo {year} {2009})}\BibitemShut {NoStop}%
\bibitem [{\citenamefont {Schmid}\ \emph {et~al.}(2013)\citenamefont {Schmid},
  \citenamefont {Opilik}, \citenamefont {Blum},\ and\ \citenamefont
  {Zenobi}}]{Schmid:2013}%
  \BibitemOpen
  \bibfield  {author} {\bibinfo {author} {\bibfnamefont {T.}~\bibnamefont
  {Schmid}}, \bibinfo {author} {\bibfnamefont {L.}~\bibnamefont {Opilik}},
  \bibinfo {author} {\bibfnamefont {C.}~\bibnamefont {Blum}}, \ and\ \bibinfo
  {author} {\bibfnamefont {R.}~\bibnamefont {Zenobi}},\ }\href {\doibase
  https://doi.org/10.1002/anie.201203849} {\bibfield  {journal} {\bibinfo
  {journal} {Angew. Chem., Int. Ed.}\ }\textbf {\bibinfo {volume} {52}},\
  \bibinfo {pages} {5940} (\bibinfo {year} {2013})}\BibitemShut {NoStop}%
\bibitem [{\citenamefont {Deckert-Gaudig}\ \emph {et~al.}(2017)\citenamefont
  {Deckert-Gaudig}, \citenamefont {Taguchi}, \citenamefont {Kawata},\ and\
  \citenamefont {Deckert}}]{Deckert:2017}%
  \BibitemOpen
  \bibfield  {author} {\bibinfo {author} {\bibfnamefont {T.}~\bibnamefont
  {Deckert-Gaudig}}, \bibinfo {author} {\bibfnamefont {A.}~\bibnamefont
  {Taguchi}}, \bibinfo {author} {\bibfnamefont {S.}~\bibnamefont {Kawata}}, \
  and\ \bibinfo {author} {\bibfnamefont {V.}~\bibnamefont {Deckert}},\ }\href
  {\doibase 10.1039/C7CS00209B} {\bibfield  {journal} {\bibinfo  {journal}
  {Chem. Soc. Rev.}\ }\textbf {\bibinfo {volume} {46}},\ \bibinfo {pages}
  {4077} (\bibinfo {year} {2017})}\BibitemShut {NoStop}%
\bibitem [{\citenamefont {Chen}\ \emph {et~al.}(2019)\citenamefont {Chen},
  \citenamefont {Hu}, \citenamefont {Mescall}, \citenamefont {You},
  \citenamefont {Basov}, \citenamefont {Dai},\ and\ \citenamefont
  {Liu}}]{Chen:2019}%
  \BibitemOpen
  \bibfield  {author} {\bibinfo {author} {\bibfnamefont {X.}~\bibnamefont
  {Chen}}, \bibinfo {author} {\bibfnamefont {D.}~\bibnamefont {Hu}}, \bibinfo
  {author} {\bibfnamefont {R.}~\bibnamefont {Mescall}}, \bibinfo {author}
  {\bibfnamefont {G.}~\bibnamefont {You}}, \bibinfo {author} {\bibfnamefont
  {D.~N.}\ \bibnamefont {Basov}}, \bibinfo {author} {\bibfnamefont
  {Q.}~\bibnamefont {Dai}}, \ and\ \bibinfo {author} {\bibfnamefont
  {M.}~\bibnamefont {Liu}},\ }\href {\doibase
  https://doi.org/10.1002/adma.201804774} {\bibfield  {journal} {\bibinfo
  {journal} {Adv. Mater.}\ }\textbf {\bibinfo {volume} {31}},\ \bibinfo {pages}
  {1804774} (\bibinfo {year} {2019})}\BibitemShut {NoStop}%
\bibitem [{\citenamefont {Wunsch}\ \emph {et~al.}(2006)\citenamefont {Wunsch},
  \citenamefont {Stauber}, \citenamefont {Sols},\ and\ \citenamefont
  {Guinea}}]{Wunsch:2006}%
  \BibitemOpen
  \bibfield  {author} {\bibinfo {author} {\bibfnamefont {B.}~\bibnamefont
  {Wunsch}}, \bibinfo {author} {\bibfnamefont {T.}~\bibnamefont {Stauber}},
  \bibinfo {author} {\bibfnamefont {F.}~\bibnamefont {Sols}}, \ and\ \bibinfo
  {author} {\bibfnamefont {F.}~\bibnamefont {Guinea}},\ }\href {\doibase
  10.1088/1367-2630/8/12/318} {\bibfield  {journal} {\bibinfo  {journal} {New
  J. Phys.}\ }\textbf {\bibinfo {volume} {8}},\ \bibinfo {pages} {318}
  (\bibinfo {year} {2006})}\BibitemShut {NoStop}%
\bibitem [{\citenamefont {Jessen}\ \emph {et~al.}(2019)\citenamefont {Jessen},
  \citenamefont {Gammelgaard}, \citenamefont {Thomsen}, \citenamefont
  {Mackenzie}, \citenamefont {Thomsen}, \citenamefont {Caridad}, \citenamefont
  {Duegaard}, \citenamefont {Watanabe}, \citenamefont {Taniguchi},
  \citenamefont {Booth}, \citenamefont {Pedersen}, \citenamefont {Jauho},\ and\
  \citenamefont {Bøggild}}]{Jessen:2019}%
  \BibitemOpen
  \bibfield  {author} {\bibinfo {author} {\bibfnamefont {B.~S.}\ \bibnamefont
  {Jessen}}, \bibinfo {author} {\bibfnamefont {L.}~\bibnamefont {Gammelgaard}},
  \bibinfo {author} {\bibfnamefont {M.~R.}\ \bibnamefont {Thomsen}}, \bibinfo
  {author} {\bibfnamefont {D.~M.~A.}\ \bibnamefont {Mackenzie}}, \bibinfo
  {author} {\bibfnamefont {J.~D.}\ \bibnamefont {Thomsen}}, \bibinfo {author}
  {\bibfnamefont {J.~M.}\ \bibnamefont {Caridad}}, \bibinfo {author}
  {\bibfnamefont {E.}~\bibnamefont {Duegaard}}, \bibinfo {author}
  {\bibfnamefont {K.}~\bibnamefont {Watanabe}}, \bibinfo {author}
  {\bibfnamefont {T.}~\bibnamefont {Taniguchi}}, \bibinfo {author}
  {\bibfnamefont {T.~J.}\ \bibnamefont {Booth}}, \bibinfo {author}
  {\bibfnamefont {T.~G.}\ \bibnamefont {Pedersen}}, \bibinfo {author}
  {\bibfnamefont {A.-P.}\ \bibnamefont {Jauho}}, \ and\ \bibinfo {author}
  {\bibfnamefont {P.}~\bibnamefont {Bøggild}},\ }\href {\doibase
  10.1038/s41565-019-0376-3} {\bibfield  {journal} {\bibinfo  {journal} {Nat.
  Nanotechnol.}\ }\textbf {\bibinfo {volume} {14}},\ \bibinfo {pages} {340}
  (\bibinfo {year} {2019})}\BibitemShut {NoStop}%
\bibitem [{\citenamefont {Rizzo}\ \emph {et~al.}(2020)\citenamefont {Rizzo},
  \citenamefont {Jessen}, \citenamefont {Sun}, \citenamefont {Ruta},
  \citenamefont {Zhang}, \citenamefont {Yan}, \citenamefont {Xian},
  \citenamefont {McLeod}, \citenamefont {Berkowitz}, \citenamefont {Watanabe},
  \citenamefont {Taniguchi}, \citenamefont {Nagler}, \citenamefont {Mandrus},
  \citenamefont {Rubio}, \citenamefont {Fogler}, \citenamefont {Millis},
  \citenamefont {Hone}, \citenamefont {Dean},\ and\ \citenamefont
  {Basov}}]{Rizzo:2020}%
  \BibitemOpen
  \bibfield  {author} {\bibinfo {author} {\bibfnamefont {D.~J.}\ \bibnamefont
  {Rizzo}}, \bibinfo {author} {\bibfnamefont {B.~S.}\ \bibnamefont {Jessen}},
  \bibinfo {author} {\bibfnamefont {Z.}~\bibnamefont {Sun}}, \bibinfo {author}
  {\bibfnamefont {F.~L.}\ \bibnamefont {Ruta}}, \bibinfo {author}
  {\bibfnamefont {J.}~\bibnamefont {Zhang}}, \bibinfo {author} {\bibfnamefont
  {J.-Q.}\ \bibnamefont {Yan}}, \bibinfo {author} {\bibfnamefont
  {L.}~\bibnamefont {Xian}}, \bibinfo {author} {\bibfnamefont {A.~S.}\
  \bibnamefont {McLeod}}, \bibinfo {author} {\bibfnamefont {M.~E.}\
  \bibnamefont {Berkowitz}}, \bibinfo {author} {\bibfnamefont {K.}~\bibnamefont
  {Watanabe}}, \bibinfo {author} {\bibfnamefont {T.}~\bibnamefont {Taniguchi}},
  \bibinfo {author} {\bibfnamefont {S.~E.}\ \bibnamefont {Nagler}}, \bibinfo
  {author} {\bibfnamefont {D.~G.}\ \bibnamefont {Mandrus}}, \bibinfo {author}
  {\bibfnamefont {A.}~\bibnamefont {Rubio}}, \bibinfo {author} {\bibfnamefont
  {M.~M.}\ \bibnamefont {Fogler}}, \bibinfo {author} {\bibfnamefont {A.~J.}\
  \bibnamefont {Millis}}, \bibinfo {author} {\bibfnamefont {J.~C.}\
  \bibnamefont {Hone}}, \bibinfo {author} {\bibfnamefont {C.~R.}\ \bibnamefont
  {Dean}}, \ and\ \bibinfo {author} {\bibfnamefont {D.~N.}\ \bibnamefont
  {Basov}},\ }\href {\doibase 10.1021/acs.nanolett.0c03466} {\bibfield
  {journal} {\bibinfo  {journal} {Nano Lett.}\ }\textbf {\bibinfo {volume}
  {20}},\ \bibinfo {pages} {8438} (\bibinfo {year} {2020})}\BibitemShut
  {NoStop}%
\bibitem [{\citenamefont {Sauls}(2018)}]{Sauls:2018}%
  \BibitemOpen
  \bibfield  {author} {\bibinfo {author} {\bibfnamefont {J.~A.}\ \bibnamefont
  {Sauls}},\ }\href {\doibase 10.1098/rsta.2018.0140} {\bibfield  {journal}
  {\bibinfo  {journal} {Philos. Trans. R. Soc. A}\ }\textbf {\bibinfo {volume}
  {376}},\ \bibinfo {pages} {20180140} (\bibinfo {year} {2018})}\BibitemShut
  {NoStop}%
\bibitem [{\citenamefont {Dienst}\ \emph {et~al.}(2013)\citenamefont {Dienst},
  \citenamefont {Casandruc}, \citenamefont {Fausti}, \citenamefont {Zhang},
  \citenamefont {Eckstein}, \citenamefont {Hoffmann}, \citenamefont {Khanna},
  \citenamefont {Dean}, \citenamefont {Gensch}, \citenamefont {Winnerl},
  \citenamefont {Seidel}, \citenamefont {Pyon}, \citenamefont {Takayama},
  \citenamefont {Takagi},\ and\ \citenamefont {Cavalleri}}]{Dienst:2013}%
  \BibitemOpen
  \bibfield  {author} {\bibinfo {author} {\bibfnamefont {A.}~\bibnamefont
  {Dienst}}, \bibinfo {author} {\bibfnamefont {E.}~\bibnamefont {Casandruc}},
  \bibinfo {author} {\bibfnamefont {D.}~\bibnamefont {Fausti}}, \bibinfo
  {author} {\bibfnamefont {L.}~\bibnamefont {Zhang}}, \bibinfo {author}
  {\bibfnamefont {M.}~\bibnamefont {Eckstein}}, \bibinfo {author}
  {\bibfnamefont {M.}~\bibnamefont {Hoffmann}}, \bibinfo {author}
  {\bibfnamefont {V.}~\bibnamefont {Khanna}}, \bibinfo {author} {\bibfnamefont
  {N.}~\bibnamefont {Dean}}, \bibinfo {author} {\bibfnamefont {M.}~\bibnamefont
  {Gensch}}, \bibinfo {author} {\bibfnamefont {S.}~\bibnamefont {Winnerl}},
  \bibinfo {author} {\bibfnamefont {W.}~\bibnamefont {Seidel}}, \bibinfo
  {author} {\bibfnamefont {S.}~\bibnamefont {Pyon}}, \bibinfo {author}
  {\bibfnamefont {T.}~\bibnamefont {Takayama}}, \bibinfo {author}
  {\bibfnamefont {H.}~\bibnamefont {Takagi}}, \ and\ \bibinfo {author}
  {\bibfnamefont {A.}~\bibnamefont {Cavalleri}},\ }\href {\doibase
  10.1038/nmat3580} {\bibfield  {journal} {\bibinfo  {journal} {Nat. Mater.}\
  }\textbf {\bibinfo {volume} {12}},\ \bibinfo {pages} {535} (\bibinfo {year}
  {2013})}\BibitemShut {NoStop}%
\bibitem [{\citenamefont {Laplace}\ and\ \citenamefont
  {Cavalleri}(2016)}]{Laplace:2016}%
  \BibitemOpen
  \bibfield  {author} {\bibinfo {author} {\bibfnamefont {Y.}~\bibnamefont
  {Laplace}}\ and\ \bibinfo {author} {\bibfnamefont {A.}~\bibnamefont
  {Cavalleri}},\ }\href {\doibase 10.1080/23746149.2016.1212671} {\bibfield
  {journal} {\bibinfo  {journal} {Adv. Phys. X}\ }\textbf {\bibinfo {volume}
  {1}},\ \bibinfo {pages} {387} (\bibinfo {year} {2016})}\BibitemShut {NoStop}%
\bibitem [{\citenamefont {Yang}\ and\ \citenamefont {Wu}(2018)}]{Yang2018}%
  \BibitemOpen
  \bibfield  {author} {\bibinfo {author} {\bibfnamefont {F.}~\bibnamefont
  {Yang}}\ and\ \bibinfo {author} {\bibfnamefont {M.~W.}\ \bibnamefont {Wu}},\
  }\href {\doibase 10.1103/PhysRevB.98.094507} {\bibfield  {journal} {\bibinfo
  {journal} {Phys. Rev. B}\ }\textbf {\bibinfo {volume} {98}},\ \bibinfo
  {pages} {094507} (\bibinfo {year} {2018})}\BibitemShut {NoStop}%
\bibitem [{\citenamefont {Tsuchiya}\ \emph {et~al.}(2018)\citenamefont
  {Tsuchiya}, \citenamefont {Yamamoto}, \citenamefont {Yoshii},\ and\
  \citenamefont {Nitta}}]{Tsuchiya2018}%
  \BibitemOpen
  \bibfield  {author} {\bibinfo {author} {\bibfnamefont {S.}~\bibnamefont
  {Tsuchiya}}, \bibinfo {author} {\bibfnamefont {D.}~\bibnamefont {Yamamoto}},
  \bibinfo {author} {\bibfnamefont {R.}~\bibnamefont {Yoshii}}, \ and\ \bibinfo
  {author} {\bibfnamefont {M.}~\bibnamefont {Nitta}},\ }\href {\doibase
  10.1103/PhysRevB.98.094503} {\bibfield  {journal} {\bibinfo  {journal} {Phys.
  Rev. B}\ }\textbf {\bibinfo {volume} {98}},\ \bibinfo {pages} {094503}
  (\bibinfo {year} {2018})}\BibitemShut {NoStop}%
\bibitem [{\citenamefont {Yu}\ and\ \citenamefont
  {Wu}(2017{\natexlab{a}})}]{Yu2017a}%
  \BibitemOpen
  \bibfield  {author} {\bibinfo {author} {\bibfnamefont {T.}~\bibnamefont
  {Yu}}\ and\ \bibinfo {author} {\bibfnamefont {M.~W.}\ \bibnamefont {Wu}},\
  }\href {\doibase 10.1103/PhysRevB.96.155311} {\bibfield  {journal} {\bibinfo
  {journal} {Phys. Rev. B}\ }\textbf {\bibinfo {volume} {96}},\ \bibinfo
  {pages} {155311} (\bibinfo {year} {2017}{\natexlab{a}})}\BibitemShut
  {NoStop}%
\bibitem [{\citenamefont {Yu}\ and\ \citenamefont
  {Wu}(2017{\natexlab{b}})}]{Yu2017b}%
  \BibitemOpen
  \bibfield  {author} {\bibinfo {author} {\bibfnamefont {T.}~\bibnamefont
  {Yu}}\ and\ \bibinfo {author} {\bibfnamefont {M.~W.}\ \bibnamefont {Wu}},\
  }\href {\doibase 10.1103/PhysRevB.96.155312} {\bibfield  {journal} {\bibinfo
  {journal} {Phys. Rev. B}\ }\textbf {\bibinfo {volume} {96}},\ \bibinfo
  {pages} {155312} (\bibinfo {year} {2017}{\natexlab{b}})}\BibitemShut
  {NoStop}%
\bibitem [{\citenamefont {Sherman}\ \emph {et~al.}(2015)\citenamefont
  {Sherman}, \citenamefont {Pracht}, \citenamefont {Gorshunov}, \citenamefont
  {Poran}, \citenamefont {Jesudasan}, \citenamefont {Chand}, \citenamefont
  {Raychaudhuri}, \citenamefont {Swanson}, \citenamefont {Trivedi},
  \citenamefont {Auerbach}, \citenamefont {Scheffler}, \citenamefont
  {Frydman},\ and\ \citenamefont {Dressel}}]{Sherman2015}%
  \BibitemOpen
  \bibfield  {author} {\bibinfo {author} {\bibfnamefont {D.}~\bibnamefont
  {Sherman}}, \bibinfo {author} {\bibfnamefont {U.~S.}\ \bibnamefont {Pracht}},
  \bibinfo {author} {\bibfnamefont {B.}~\bibnamefont {Gorshunov}}, \bibinfo
  {author} {\bibfnamefont {S.}~\bibnamefont {Poran}}, \bibinfo {author}
  {\bibfnamefont {J.}~\bibnamefont {Jesudasan}}, \bibinfo {author}
  {\bibfnamefont {M.}~\bibnamefont {Chand}}, \bibinfo {author} {\bibfnamefont
  {P.}~\bibnamefont {Raychaudhuri}}, \bibinfo {author} {\bibfnamefont
  {M.}~\bibnamefont {Swanson}}, \bibinfo {author} {\bibfnamefont
  {N.}~\bibnamefont {Trivedi}}, \bibinfo {author} {\bibfnamefont
  {A.}~\bibnamefont {Auerbach}}, \bibinfo {author} {\bibfnamefont
  {M.}~\bibnamefont {Scheffler}}, \bibinfo {author} {\bibfnamefont
  {A.}~\bibnamefont {Frydman}}, \ and\ \bibinfo {author} {\bibfnamefont
  {M.}~\bibnamefont {Dressel}},\ }\href {\doibase 10.1038/nphys3227} {\bibfield
   {journal} {\bibinfo  {journal} {Nat. Physics}\ }\textbf {\bibinfo {volume}
  {11}},\ \bibinfo {pages} {188} (\bibinfo {year} {2015})}\BibitemShut
  {NoStop}%
\bibitem [{\citenamefont {Podolsky}\ \emph {et~al.}(2011)\citenamefont
  {Podolsky}, \citenamefont {Auerbach},\ and\ \citenamefont
  {Arovas}}]{Podolsky2011}%
  \BibitemOpen
  \bibfield  {author} {\bibinfo {author} {\bibfnamefont {D.}~\bibnamefont
  {Podolsky}}, \bibinfo {author} {\bibfnamefont {A.}~\bibnamefont {Auerbach}},
  \ and\ \bibinfo {author} {\bibfnamefont {D.~P.}\ \bibnamefont {Arovas}},\
  }\href {\doibase 10.1103/PhysRevB.84.174522} {\bibfield  {journal} {\bibinfo
  {journal} {Phys. Rev. B}\ }\textbf {\bibinfo {volume} {84}},\ \bibinfo
  {pages} {174522} (\bibinfo {year} {2011})}\BibitemShut {NoStop}%
\end{thebibliography}%

\end{document}